\documentclass[twocolumn,floatfix, showpacs]{revtex4}

\usepackage[final]{epsfig}
\usepackage{amsmath}
\usepackage{parskip}
 
\begin{document}
 
\title{Medium-modified Jet Shapes and other Jet Observables from in-medium Parton Shower Evolution}
 
\author{Thorsten Renk}
\email{trenk@phys.jyu.fi}
\affiliation{Department of Physics, P.O. Box 35 FI-40014 University of Jyv\"askyl\"a, Finland}
\affiliation{Helsinki Institute of Physics, P.O. Box 64 FI-00014, University of Helsinki, Finland}
 

\begin{abstract}
The suppression of large transverse momentum ($P_T$) hadrons in heavy-ion (A-A) collisions as compared to their scaled expectation from proton-proton (p-p) collisions due to the interaction of hard partons with the hot and dense QCD medium in A-A collisions is experimentally a well established phenomenon. Focusing on leading hadrons produced in hard processes, the medium effect appears as energy loss. Beyond that, the question is how the lost energy is redistributed in the medium. With increased experimental statistics and most importantly the kinematic range of the LHC, studying the properties of full jets rather than leading hadrons is becoming feasible. On the theory side, analytic models and Monte-Carlo (MC) codes for in-medium shower evolution are being developed to describe jets in the medium. In this paper, expectations for medium-modified jet observables, the jet shapes, the thrust distribution and the $n$-jet fraction, are computed with the MC code YaJEM for various scenarios of the parton-medium interaction which all are consistent with high $P_T$ hadron suppression data. The computation is done at 20 and 100 GeV jet energy, corresponding to probing typical RHIC and LHC kinematics, and the possibility to make an unbiased measurement of the observables is discussed.
\end{abstract}
 
\maketitle

\section{Introduction}

Jet quenching, i.e.\ the interaction of hard partons created in the first moments of a heavy ion collision with the surrounding medium of strongly interacting matter has long been regarded a promising tool to study properties of that medium \cite{Jet1,Jet2,Jet3,Jet4,Jet5,Jet6}. The guiding principle is to study the changes to a hard process well-known in p-p collisions which are induced by the medium. A number of observables which essentially probe the modification of the leading hadron in a jet are available for this purpose, among them suppression of single inclusive hard hadron spectra in terms of the nuclear suppression factor $R_{AA}$ \cite{PHENIX_R_AA}, the suppression of back-to-back correlations \cite{Dijets1,Dijets2} and also single hadron suppression as a function of the emission angle with the reaction plane \cite{PHENIX-RP}. Most recently also preliminary measurements of fully reconstructed jets have become available \cite{STARJET}.

Single hadron observables and back-to-back correlations are well described in detailed model calculations using the concept of energy loss \cite{HydroJet1,Dihadron1,Dihadron2}, i.e. under the assumption that the process can be described by a medium-induced shift of the leading parton energy by an amount $\Delta E$, followed by a fragmentation process using vacuum fragmentation of a parton with the reduced energy. However, there are also calculations for these observables in which the evolution of the whole in-medium parton shower is followed in an analytic way \cite{HydroJet2,HydroJet3,Dihadron3}. Recently, also Monte Carlo (MC) codes for in-medium shower evolution have become available \cite{JEWEL,YAS,YAS2,YAS3,Carlos} which are based on MC shower simulations developed for $e^+e^-$ or p-p collisions, such as PYTHIA \cite{PYTHIA} or HERWIG \cite{HERWIG}, in which the shower evolves in vacuum. In a medium-modified shower evolution, in contrast to energy loss calculations, energy is not simply lost but redistributed in a characteristic way. 

Currently the precise nature of the jet-medium interaction is not known. In \cite{YAS3} we have studied three different scenarios including medium-induced radiation and a drag force in their effect on various observables in the RHIC kinematic range using the im-medium shower MC code YaJEM (Yet another Jet Energy-loss Model). The aim of the present paper is to extend this study to the LHC range and to discuss other observables which probe the medium-induced redistribution of energy and are hence sensitive to the nature of the parton-medium interaction.

Observables such as jet shapes allow to study the energy redistribution caused by the medium e.g. as a modification of the angular flow of energy inside the jet. Thus, in this paper we mainly focus on jet shapes, i.e. the transverse flow profile of the jet energy, and their in-medium modification. Analytical results for this observable have already been obtained previously \cite{Shapes1,Shapes2}. In the medium, there is always an ambiguity if a given parton is considered to be part of the jet or part of the medium. Momentum and angular cuts may provide a separation criterion, however in turn introduce a bias which affects any given jet observable. Thus, we also study the effect of such cuts on the jet shape. In addition, we compute other jet observables such as the thrust distribution or the $n-$jet fraction which have been used in $e^+e^-$ collisions for precision tests of QCD \cite{ALEPH}. While these quantities may not be easy to observe in heavy-ion collisions due to the background medium and the presence of multiple hard processes in the same event, the comparison with the well established vacuum results and the in-medium results obtained with the different MC code JEWEL \cite{JEWEL} is nevertheless instructive.

\section{The model}

The MC code YaJEM is described in great detail in \cite{YAS,YAS3}, therefore we will only give a brief outline here. The aim of the code is to compute the evolution of a QCD parton shower following a hard process in the medium. In the absence of a medium, the evolution of the shower is computed using the PYSHOW routine \cite{PYSHOW} which is part of the PYTHIA package \cite{PYTHIA}.

In the medium, the main assumption of YaJEM is that the kinematics or the branching probability of a parton are altered during its propagation. For this purpose, it is necessary to make a link between the shower evolution which is computed in momentum space and the evolution of the medium which is computed in position space. We assume that the average formation time of a shower parton with virtuality $Q$ is developed on the timescale $1/Q$, i.e. the average lifetime of a virtual parton $b$ with virtuality $Q_b$ and energy $E_b$ coming from a parent parton with virtuality $Q_a$ is in the rest frame of the original hard collision (the local rest frame of the medium may be different by a flow boost as the medium may not be static) given by

\begin{equation}
\label{E-Lifetime}
\langle \tau_b \rangle = \frac{E_b}{Q_b^2} - \frac{E_b}{Q_a^2}.
\end{equation}  

Going beyond the ansatz of \cite{YAS,YAS2} where we used this average formation time for all partons we assume as in \cite{YAS3} that the actual formation time $\tau_b$ can be obtained from a probability distribution 

\begin{equation}
\label{E-RLifetime}
P(\tau_b) = \exp\left[- \frac{\tau_b}{\langle \tau_b \rangle}  \right]
\end{equation}

which we sample to determine the actual formation time of the fluctuation in each branching. This establishes the temporal structure of the shower. With regard to the spatial structure in terms of spacetime rapidity $\eta_s$, transverse radius $r$ and angle $\phi$ , we make the simplifying assumption that all partons probe the medium along the eikonal trajectory of the shower initiating parton, i.e. we neglect the small difference of the velocity of massive partons to the speed of light and possible (equally small) changes of medium properties within the spread of the shower partons transverse to the shower axis. 

YaJEM provides three different scenarios for the effect of the medium on the evolving shower. In the RAD scenario, the idea is that the virtuality of a shower parton can grow during its propagation through the medium. Such increased virtuality subsequently leads to additional medium-induced branching.  In practice, we increase the virtuality of a shower parton $a$ propagating through a medium with specified transport coefficient $\hat{q}(\eta_s, r ,\phi,\tau)$ by

\begin{equation}
\label{E-Qgain}
\Delta Q_a^2 = \int_{\tau_a^0}^{\tau_a^0 + \tau_a} d\zeta \hat{q}(\zeta)
\end{equation}

where the time $\tau_a$ is given by Eq.~(\ref{E-RLifetime}), the time $\tau_a^0$ is known in the simulation as the endpoint of the previous branching process and the integration $d\zeta$ is along the eikonal trajectory of the shower-initiating parton. If the parton is a gluon, the virtuality transfer from the medium is increased by the ratio of their Casimir color factors, $3/\frac{4}{3} =  2.25$.

In the DRAG scenario which is motivated by results from modelling QCD-like $N=4$ super Yang-Mills theories via the AdS/CFT conjecture \cite{AdS}, we assume that the medium exerts a drag force on each propagating parton. The medium is thus characterized by a drag coefficient $D(\eta_s, r ,\phi,\tau)$ which describes the energy loss per unit pathlength. In the simulation, the energy (and momentum) of a parton $a$ are reduced by

\begin{equation}
\label{E-Drag}
\Delta E_a = \int_{\tau_a^0}^{\tau_a^0 + \tau_a} d\zeta D(\zeta).
\end{equation}

The third scenario has been suggested in \cite{JEWEL,HBP} and is included here for comparison. In the following, it is referred to as FMED. Here, the modification does not concern the parton kinematics, but rather the branching probabilities. In this scenario, the singular part of the branching kernel in the medium is enhanced by a factor $(1+f_{med})$ where $f_{med}$ represents an average measure of the medium effect for a given path through the medium.

We assume that all the three relevant parameters $\hat{q},D,f_{med}$ can be linked with the medium energy density up to a (possibly dimensionful) constant factor as

\begin{equation}
\hat{q},D \sim K  [\epsilon(\zeta)]^{3/4} (\cosh \rho(\zeta) - \sinh \rho(\zeta) \cos\psi).
\end{equation}

and 

\begin{equation}
\label{E-fmed}
f_{med} \sim K \int d \zeta [\epsilon(\zeta)]^{3/4} (\cosh \rho(\zeta) - \sinh \rho(\zeta) \cos\psi).
\end{equation}

Here, $\rho$ is the transverse flow of the medium at position $\zeta$ and $\psi$ is the angle between flow and the direction of parton propagation. For most possible paths of partons occuring in a realistic medium, the results in \cite{YAS,YAS3} have shown that the precise dependence of $\epsilon(\zeta)$ has a small effect on the shower evolution. In essence paths through an expanding medium can be considered equivalent if the line integral along the eikonal path of the shower initiating parton results in the same value of $\Delta Q^2$ ($\Delta E$). We will use this scaling law extensively in the following (note that it is by definition fulfilled in the FMED scenario). 

The model assumptions presented here with regard to the medium effect on the shower are (for the moment) rather qualitative. Complementary pQCD approaches to the medium effect such as the high order opacity expansion which can also be solved in a MC scheme \cite{Opacity} are more focused on treating the quantum interference between subsequent scatterings in the medium correctly and can hence be used to test the model assumptions made here, or ultimately be envisioned as an extension of the present work.

\section{Separation of jet and medium}

For a shower evolving in the medium, there is in general exchange of energy and momentum between what one considers shower partons and what one considers the medium. Note that within the framework of YaJEM, in the RAD scenario the shower {\em gains} energy from the medium by means of the virtuality increase, in the DRAG scenario the shower {\em loses} energy to the medium whereas the shower energy is conserved in the FMED scenario. While this appears surprising at first, it is actually rather a matter of book-keeping and defining a distinction between shower and medium.

There is no conceptual way to actually distinguish soft shower partons from soft medium partons. Within the model framework, an artificial distinction is maintained since all partons which explicitly appear are created in branchings and hence are considered to be part of the shower while the medium appears never formulated in terms of partons, but only as an effective influence on the shower. It is this artificial separation between shower and medium which makes for the surprising above results: In the RAD scenario, the possibility of soft partons being absorbed by the medium and hence transferring energy to the medium is not considered, as a result there is no energy transfer from shower into the medium and thus the shower energy can only increase. In stark contrast, in the DRAG scenario it is assumed that the medium can absorb energy without medium partons becoming in any way correlated with the shower (for example, in elastic parton-medium interactions, recoiling medium partons become correlated with the jet direction). As a result, there is no possibility to transfer energy from the medium into the shower. Finally, in the FMED scenario the somewhat artificial assumption is made that the medium can modify branching probabilities without any explicit energy-momentum transfer, hence the energy in the shower remains conserved.

In a more realistic model in which the structure of the medium is resolved, one would define a criterion (say a momentum scale) based on which partons are removed from the shower and become part of the medium. In such a model, all three scenarios would lead to a net loss of energy from the shower to the medium through the appearance of soft partons in the evolution, in addition to possible other mechanisms of energy transfer from jet to the medium. Unfortunately we do not yet have such a model for the medium, since one goal of the hard probes program is to establish the structure of the medium in terms of its relevant degrees of freedom and their properties.
However, lacking such a framework which must also be able to account fully for the back-reaction of the medium to the energy deposition by the jet to be consistent, we may still ignore the complications of the medium by imposing a momentum cutoff and focusing exclusively on the hard part of the jet for which the above caveats do not apply.

Based on hydrodynamical models of the medium, a reasonable cut to separate the perturbative dynamics of in-medium shower evolution from the soft, non-perturbative dynamics of the medium is $P_{T_{min}} = 2$ GeV at RHIC energies and $P_{T_{min}} = 4$ GeV at LHC energies. Note that such a cutoff introduces a substantial bias on the sample of jets which enter the computation of a given observable. To gauge this effect, we will in the following discuss all observables also without the cut imposed. This is done with the explicit understanding that these quantities do not correspond to observables, as they each neglect an important part of the dynamics between jet and medium for which there is no reason to be described by perturbative QCD. On the other hand, quantities where the momentum cutoff is imposed should be in principle comparable with experimental results, although in practice complications such as the experimental jet finding strategy or the presence of multiple hard processes in a single event need to be taken into account before making an actual comparison.

\section{Thrust and $n$-jet fraction}

\begin{figure*}[htb]
\begin{center}
\epsfig{file=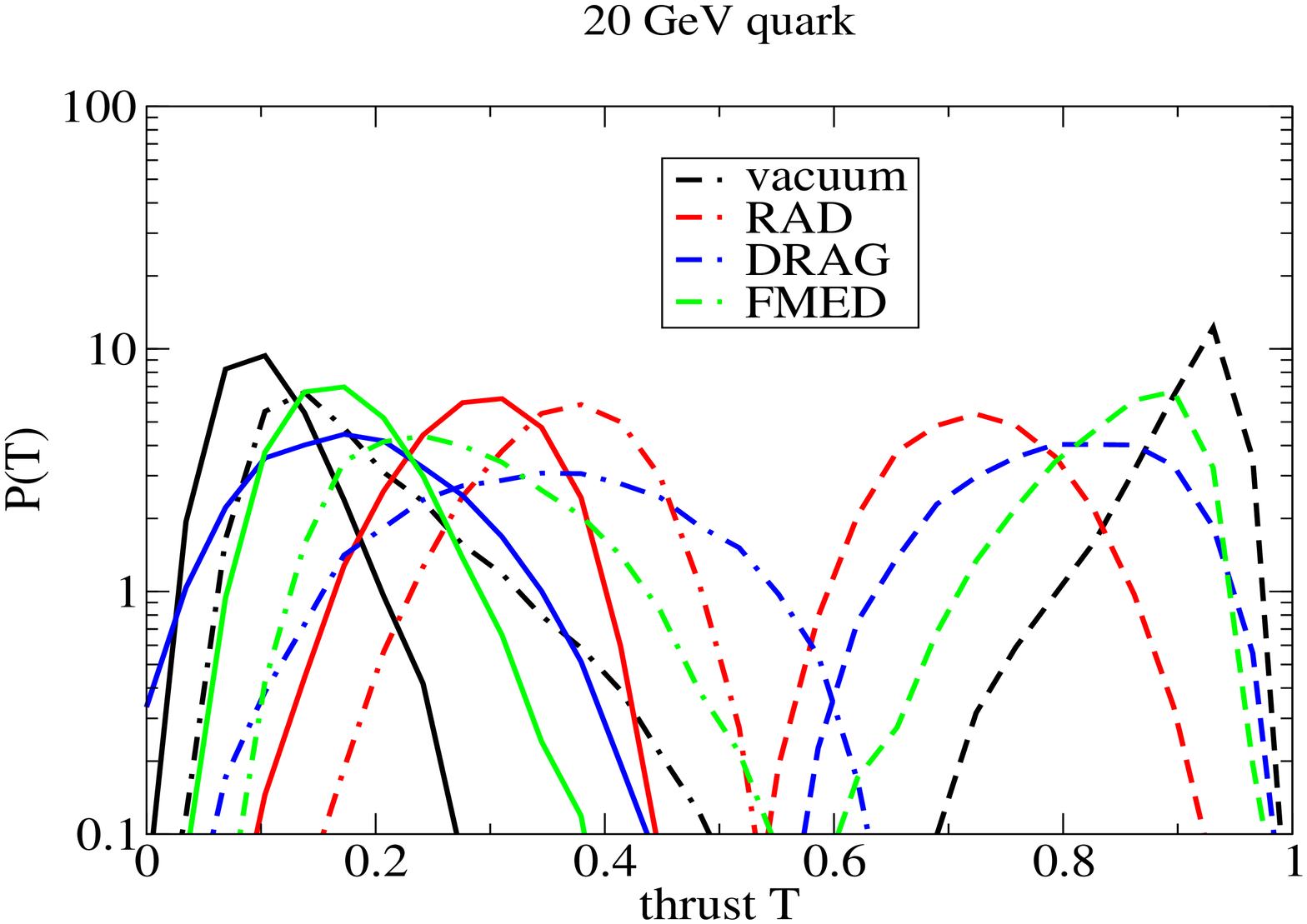,width=8cm}\epsfig{file=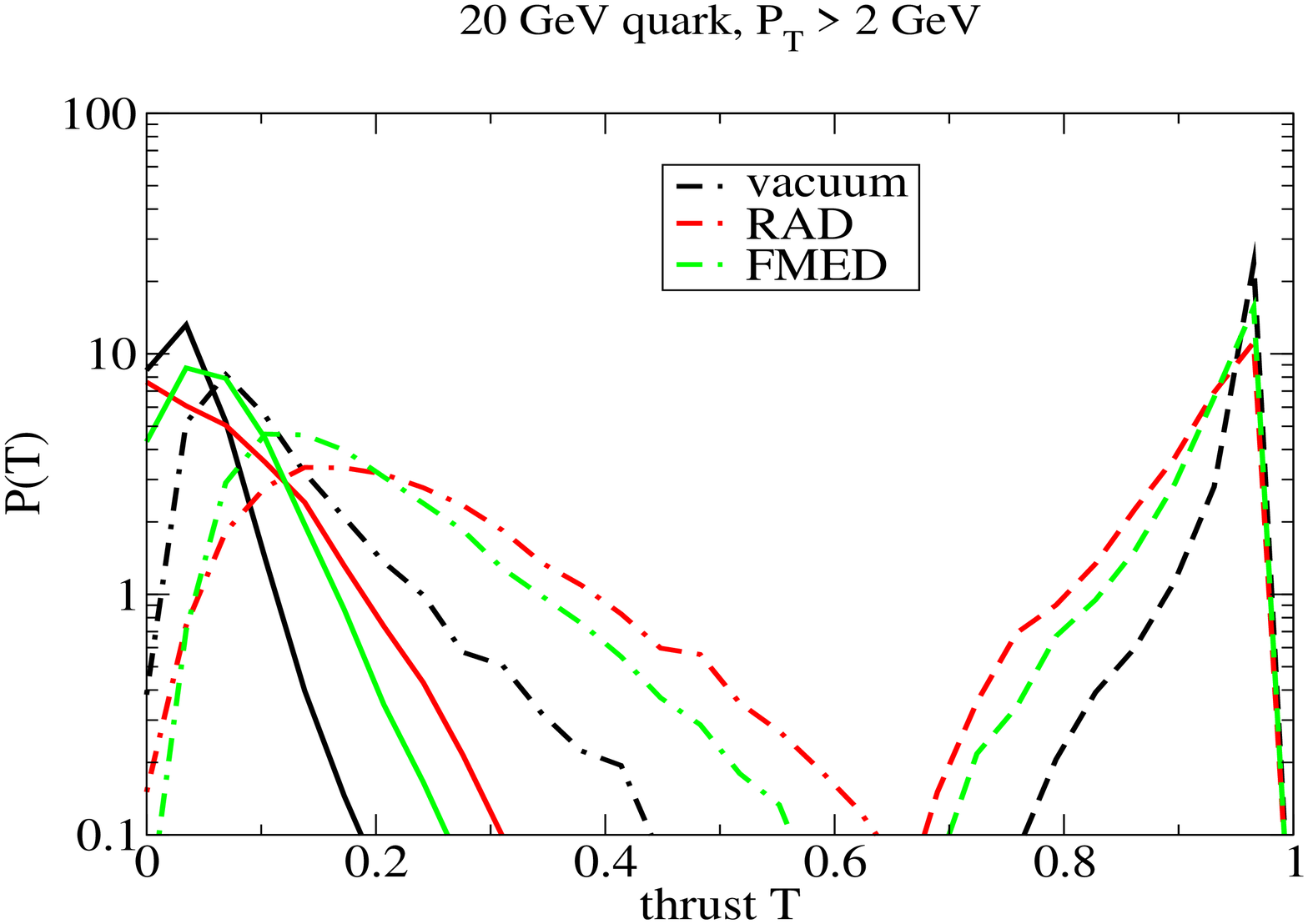, width=8cm}
\end{center}
\caption{\label{F-ThrustRHIC}(Color online) Distribution of thrust $T$ (dashed), $T_{maj}$ (solid) and $T_{min}$ (dash-dotted) for a 20 GeV shower initiating quark in vacuum and three different scenarios for parton-medium interaction (see text). Shown is the distribution for all final state particles (left panel) and for those above 2 GeV (right panel). Note that the DRAG scenario for the given choice of medium parameters renders most events below the cutoff, thus a thrust analysis above the imposed cutoff cannot be done.}
\end{figure*}

\begin{figure*}[htb]
\begin{center}
\epsfig{file=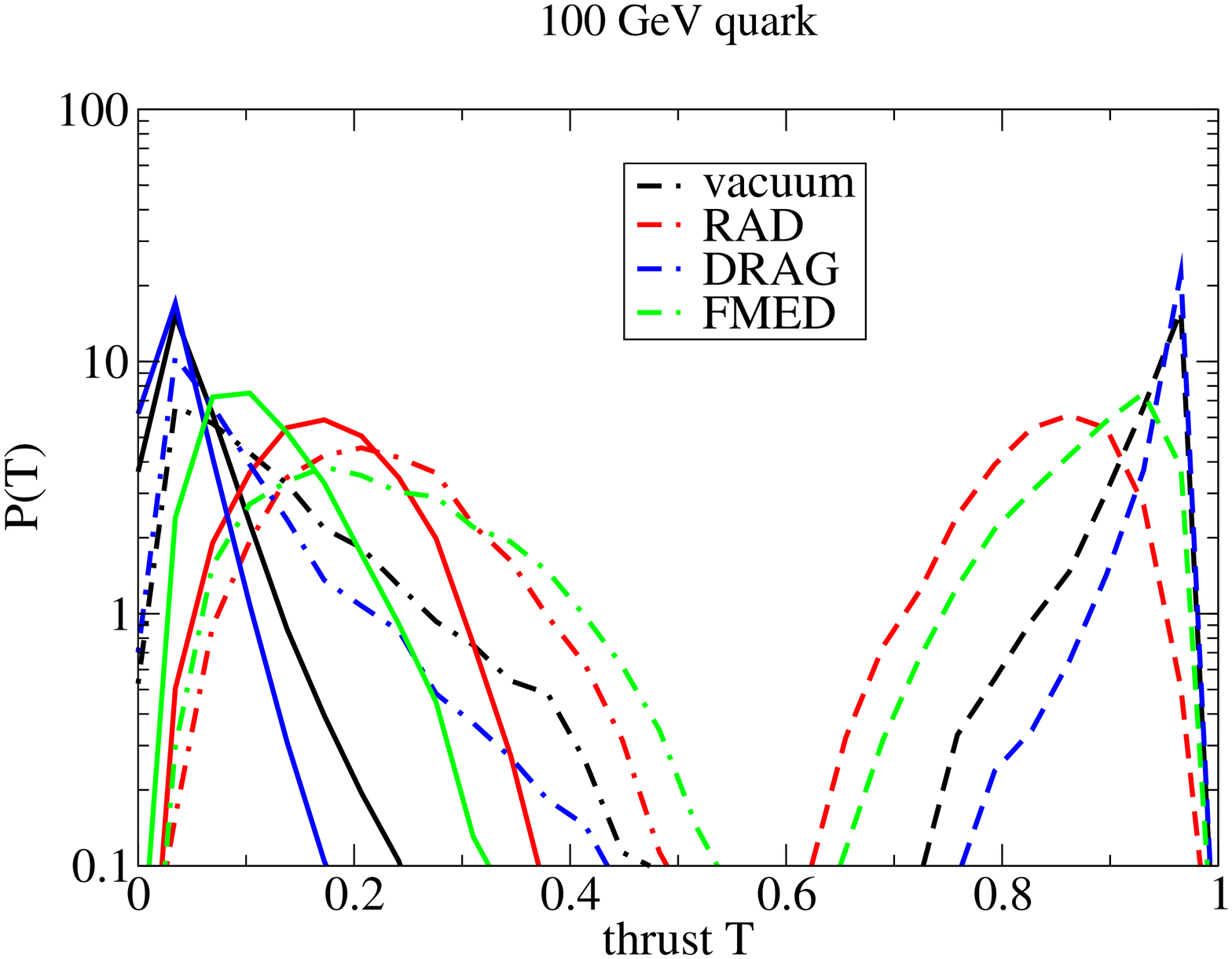,width=8cm}\epsfig{file=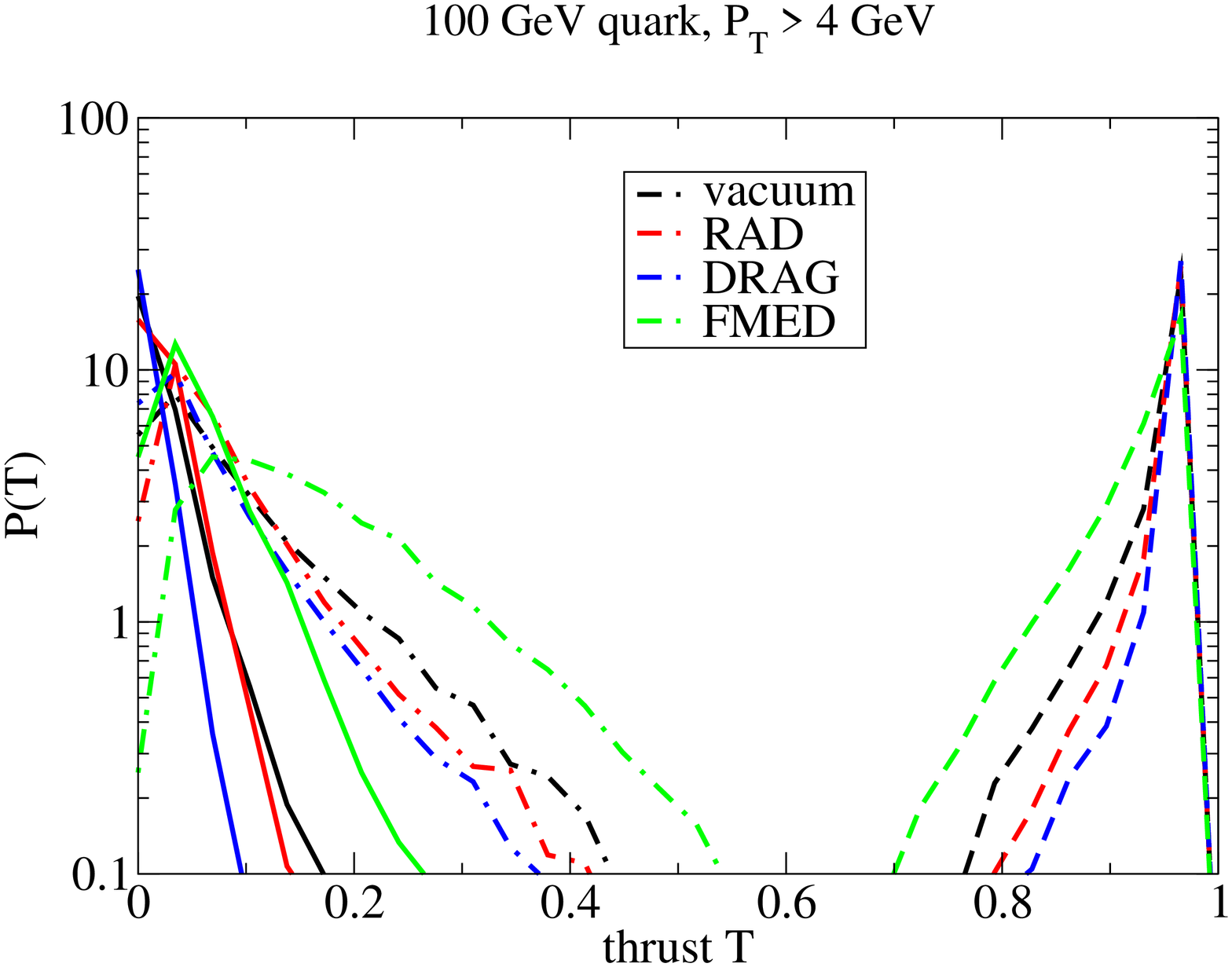, width=8cm}
\end{center}
\caption{\label{F-ThrustLHC}(Color online) Distribution of thrust $T$ (dashed), $T_{maj}$ (solid) and $T_{min}$ (dash-dotted) for a 100 GeV shower initiating quark in vacuum and three different scenarios for parton-medium interaction (see text). Shown is the distribution for all final state particles (left panel) and for those above 4 GeV (right panel). }
\end{figure*}

The overall flow of energy in an event can be traced by three event shape observables, namely thrust $T$, thrust major $T_{maj}$ and thrust minor $T_{min}$. They involve a summation over all final state particles in an event. The ALEPH collaboration has measured thrust distributions in $e^+e^-$ collisions \cite{ALEPH} in the absence of any medium-induced final state effect. In this system, PYTHIA is able to account for all observables well, thus establishing a baseline for the subsequent discussion. A detailed comparison of PYTHIA with the ALEPH data can be found in \cite{ALEPH}.

Thrust is defined as a sum over all final state particle three momenta ${\bf p}_i$

\begin{equation}
T = \text{max}_{{\bf n}_T} \frac{\sum_i | {\bf p}_i \cdot {\bf n}_T|}{\sum_i|{\bf p}_i|}.
\end{equation}

It measures how well the final state hadrons are aligned in an axis defined by the shower originating partons. For $T=1$ this alignment is perfect, for $T=0.5$ the event is spherical and no preferred axis can be identified. Thrust major is the projection of all particle momenta on this axis

\begin{equation}
T_{maj} = \text{max}_{{\bf n}_T \cdot {\bf n}=0} \frac{\sum_i | {\bf p}_i \cdot {\bf n}|}{\sum_i|{\bf p}_i|}
\end{equation}

while thrust minor sums components of momenta which are orthogonal to ${\bf n}$ and ${\bf n}_T$

\begin{equation}
T_{min} = \frac{\sum_i | {\bf p}_i \cdot {\bf n}_{mi}|}{\sum_i |{\bf p}_i|}
\end{equation}

where ${\bf n}_{mi} = {\bf n}_T \times {\bf n}$.

While the sum over all final state particles in $e^+e^-$ collisions can be traced back to a single hard process, this may not be so in A-A collisions where the number of binary collisions is $O(1000)$. Thus, even with a cut in $P_{T_{min}}$ imposed, the calculated thrust distribution may not correspond to the observable ones.

In Fig.~\ref{F-ThrustRHIC} we show the distributions of $T, T_{maj}$ and $T_{min}$ for showers initiated by a 20 GeV quark for a medium path with $\Delta Q^2_{tot}=15$ GeV$^2$, $\Delta E_{tot} = 15$ GeV and $f_{med} = 1.5$. The proportionality between these parameters is chosen based on the criterion that the medium-modified fragmentation functions computed for them approximately agree in an interval of $0.4 < z < 0.7$. This is the region of the fragmentation function which is predominantly probed when the fragmentation function is folded with a pQCD parton spectrum to compute single inclusive hadron production. In essence, for this proportionality between the parameters all scenarios reproduce the measured single hadron suppression $R_{AA}$ equally well \cite{YAS3}, and hence the results should in some sense be comparable. The particular choice of $\Delta Q^2_{tot}=15$ GeV then corresponds to a path through the whole RHIC medium, i.e. a medium modification close to the maximum which is possible.

It is evident from the figure that at this kinematics the medium tends to make the event more spherical, i.e. the distributions widen and move towards 0.5. While the effects for all particles are dramatic, the observable distribution above the imposed cutoff still shows some modification. The effect of both radiative scenarios RAD and FMED is qualitatively similar and agrees with what has been reported in \cite{JEWEL} using the MC code JEWEL.

We show the same distribution for a 100 GeV shower initiating quark  as appropriate for the LHC kinematic in Fig.~\ref{F-ThrustLHC}. For the sake of comparison, we have chosen the medium parameters $\Delta Q^2_{tot}, \Delta E_{tot}$ and $f_{med}$ identical to the RHIC case above. However, since the medium at LHC is expected to be somewhat more dense \cite{Hydro}, these values represent now a fairly typical path through the medium rather than the maximal medium effect as in the  RHIC case.

It can be noted that without the cutoff the DRAG scenario acts rather different at higher $P_T$ --- here it tends to focus the distribution towards $T=1$ whereas the radiative scenarios RAD and FMED still make the event more spherical. However, RAD does so only at low $P_T$ whereas FMED does so at all $P_T$ --- if a cutoff is imposed, it is evident that the high $P_T$ part of the thrust distribution is not widened in the RAD scenario which is a possible distinctive feature between the otherwise rather similar RAD and FMED prescriptions.

\begin{figure*}[htb]
\begin{center}
\epsfig{file=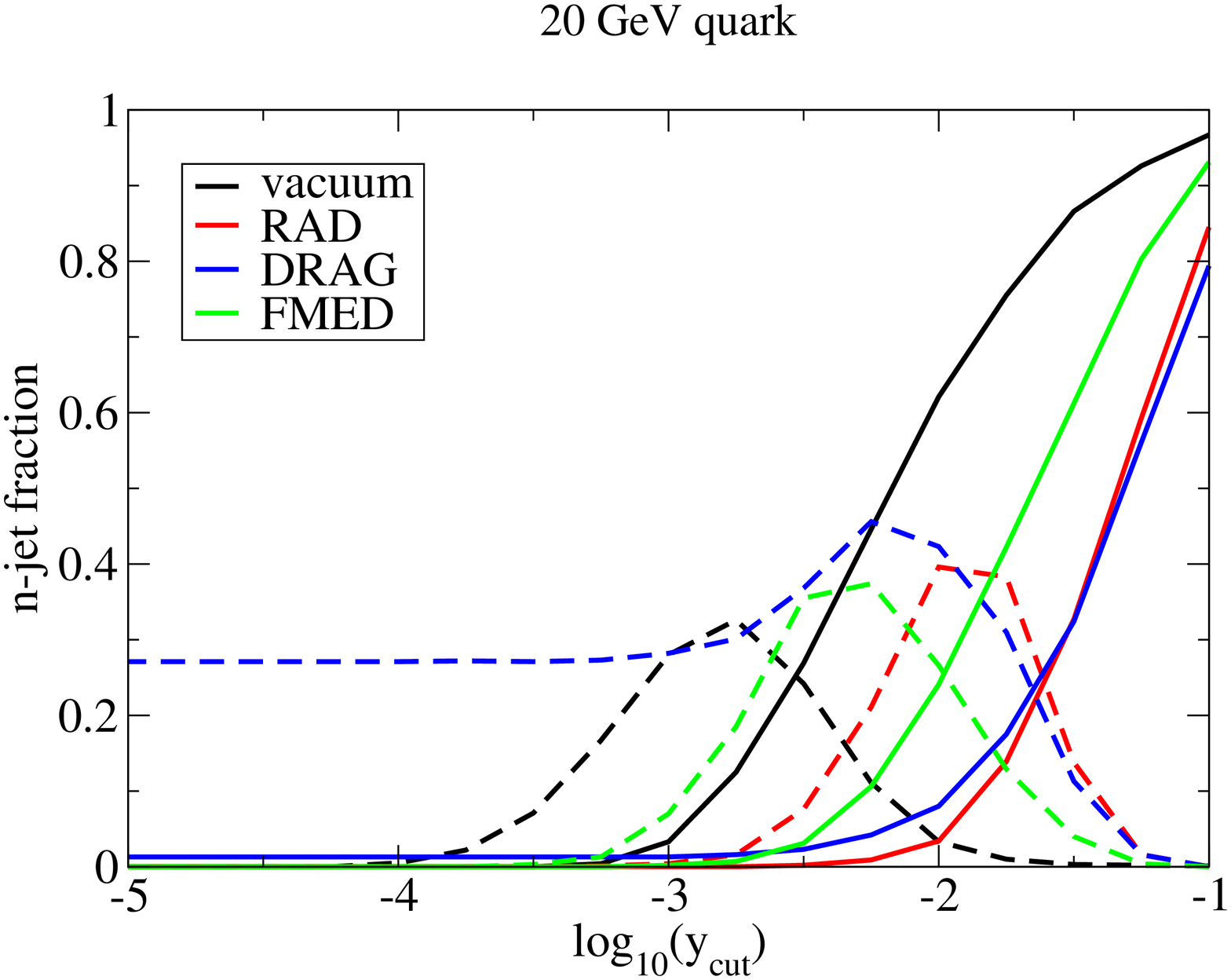,width=8cm}\epsfig{file=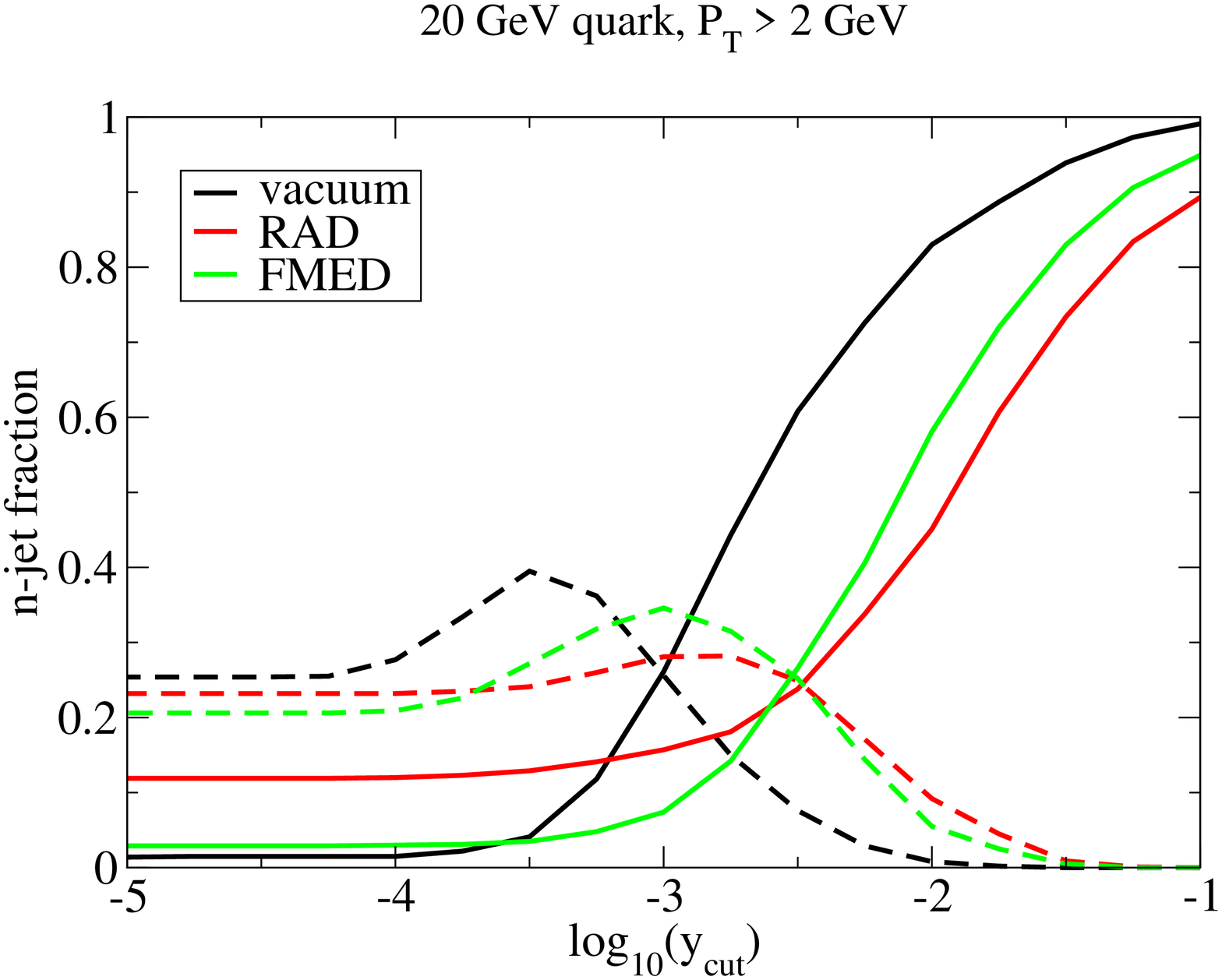, width=8cm}
\end{center}
\caption{\label{F-njet}(Color online) The two (solid) and four (dashed) jet fraction for two back-to-back shower initiating quarks with 20 GeV energy each in vacuum for three different scenarios for parton-medium interaction (see text). Shown is the distribution for all final state particles (left panel) and for those above 2 GeV (right panel). Note that the DRAG scenario for the given choice of medium parameters renders most events below the cutoff, thus a clustering analysis above the imposed cutoff cannot be done.}
\end{figure*}

\begin{figure*}[htb]
\begin{center}
\epsfig{file=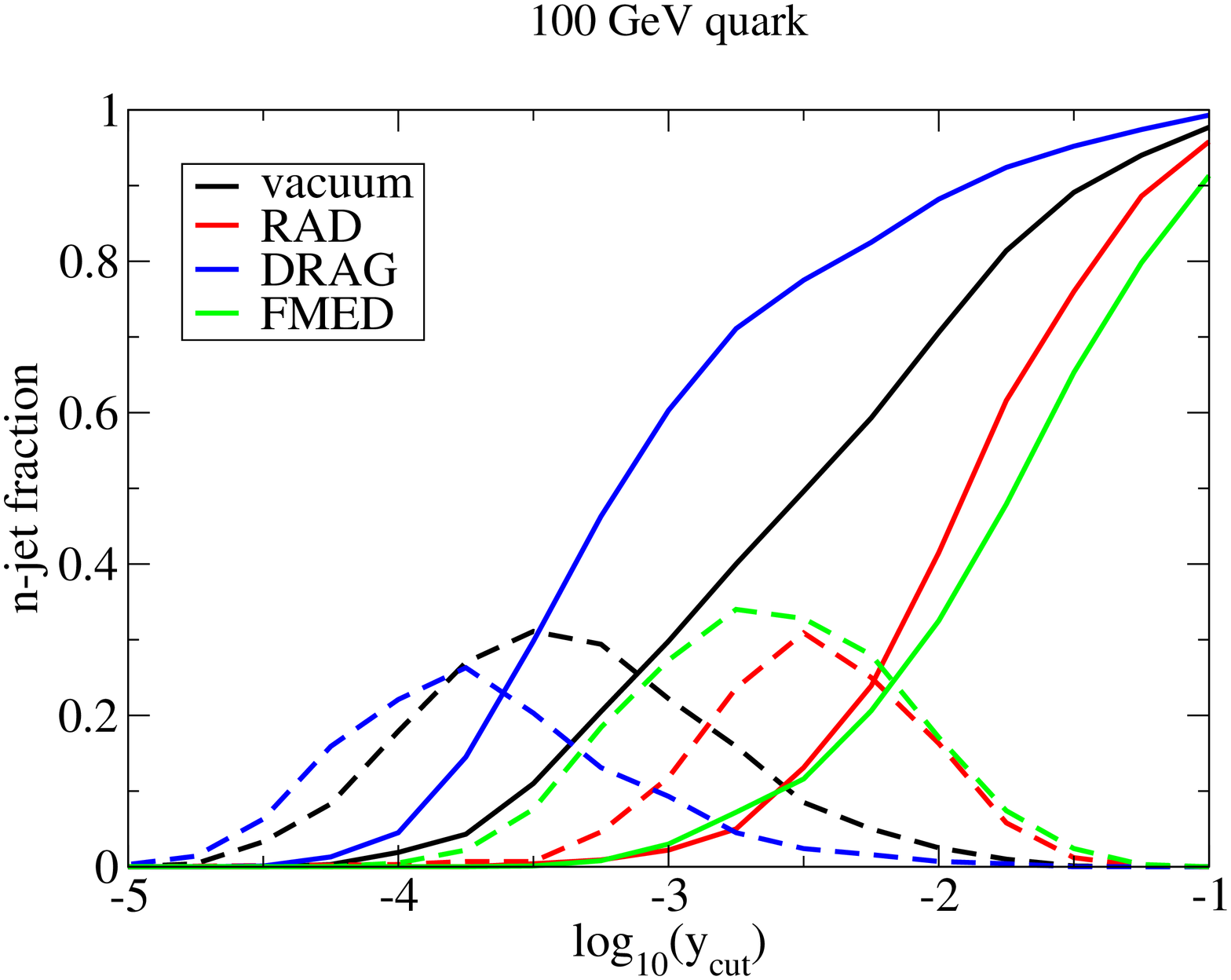,width=8cm}\epsfig{file=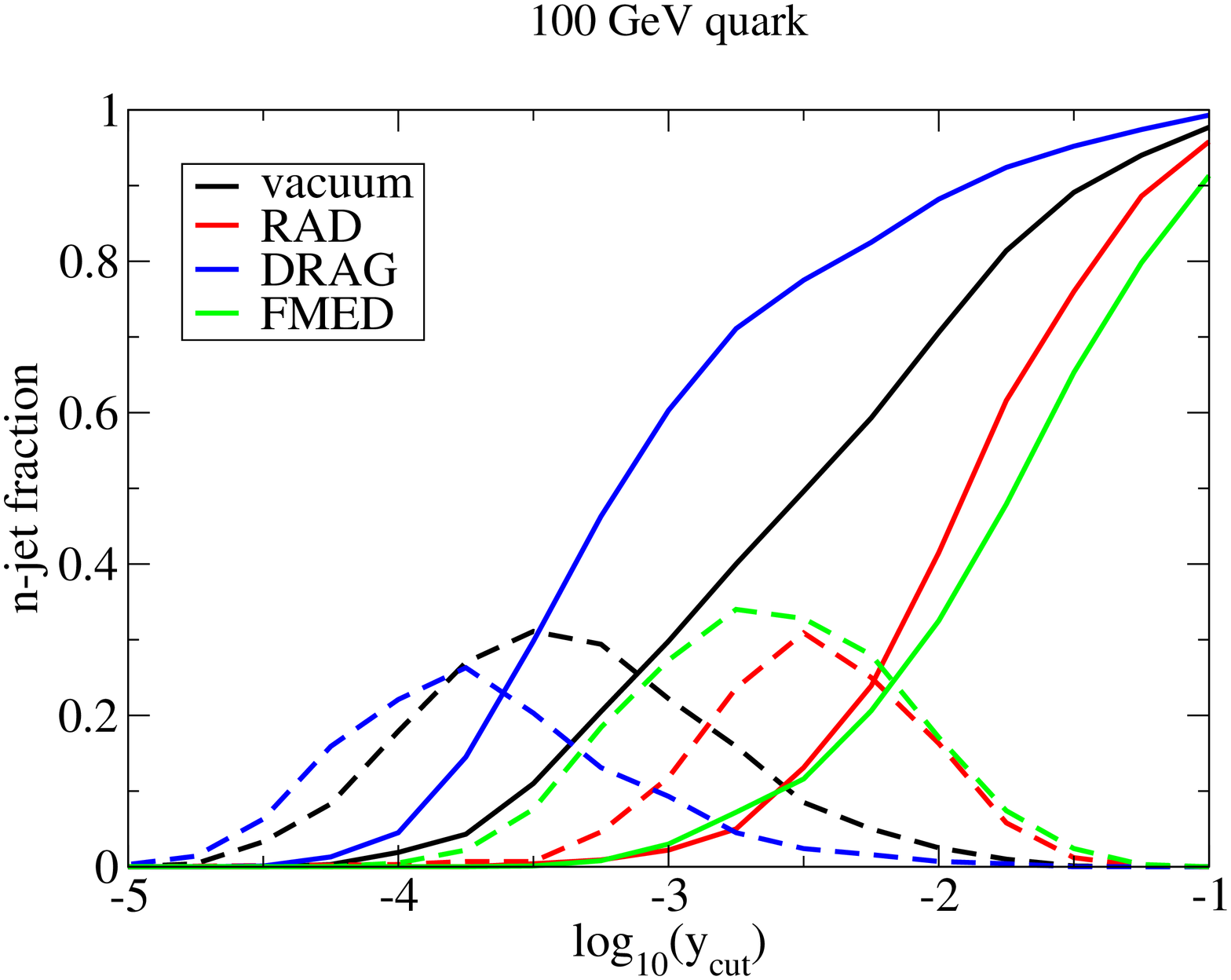, width=8cm}
\end{center}
\caption{\label{F-njetLHC}(Color online) The two (solid) and four (dashed) jet fraction for two back-to-back shower initiating quarks with 100 GeV energy each in vacuum for three different scenarios for parton-medium interaction (see text). Shown is the distribution for all final state particles (left panel) and for those above 4 GeV (right panel). }
\end{figure*}

Let us now focus on a different observable, the $n$-jet fraction. This observable is sensitive to the substructure of a jet in terms of clusters of hadrons created by the dynamics of showering and hadronization. It is based on the Durham clustering algorithm \cite{Durham}. This algorithm clusters final state particles based on a distance measure between a pair $i,j$

\begin{equation}
y_{ij} = 2 \text{min}(E_i^2,E_j^2) (1-\cos(\theta_{ij})/E_{\text{cm}}^2.
\end{equation}

This $y_{ij}$ is a measure for the transverse momentum of a softer particle of a pair with respect to the axis defined by the harder particle. In each step, the clustering algorith replaces the pair with the smallest $y_{ij}$ by a cluster with energy and momentum given by the sum of the pair's energy and momentum. The procedure is repeated and particles and clusters are further merged until $y_{ij}$ exceeds a pre-defined threshold $y_{cut}$ which corresponds to a given resolution scale. The number of clusters present at this point is called $n$ for this event, given $y_{cut}$. In averaging over many events, one finds for each choice of $y_{cut}$ a fraction of events with $n= 1,2,3, \dots$ events --- this is referred to as the $n$-jet fraction.

The same caveats which were relevant for the thrust distribution hold for the $n$-jet fraction: Even with a cutoff imposed, one cannot necessarily assume that there is only a single hard event in a heavy-ion collision, and the effect of multi-jet events must be accounted for before a comparison with data can be made.

We show the 2-jet and 4-jet fraction as a function of the resolution scale $y_{cut}$ for a back-to-back pair of 20 GeV quarks corresponding to RHIC kinematics in vacuum and in medium (with medium parameters chosen as before) in Fig.~\ref{F-njet}. For a coarse resolution ($\log_{10}(y_{cut} \sim -1$) the algorithm picks up two jets corresponding to the back-to-back event, when the resolution scale is increased, finding more clusters becomes increasingly likely. Note that some scenarios (especially with the cutoff imposed) saturate for increased resolution scale. This is caused by the relatively small multiplicity in the event, either when (as in the DRAG scenario) a large amount of energy is transferred to the medium or when a cutoff is imposed while soft particle production is enhanced. No increase in resolution scale can find six clusters if there are only four particles in the final state above the cutoff.

For RHIC kinematic conditions and with the cutoff imposed, the effect of the medium is to increase the number of clusters seen at a given resolution scale. This agrees with the previous observation based on the thrust distribution that the medium tends to make the event more spherical --- this also implies that the spread in transverse momenta is increased, and this is chiefly what is seen in the plots.

The 2-jet and 4-jet fraction for LHC kinematics, i.e. a back-to-back event of quarks with 100 GeV energy each is shown in Fig.~\ref{F-njetLHC}. Here the radiative scenarios RAD and FMED show qualitatively the same picture as for RHIC --- at the same resolution scale they tend to lead to more clusters. In contrast to this is the DRAG scenario which shows the opposite trend. This agrees with the thrust distribution in the DRAG scenario, which show that the events become more focused. Combined with the fact that overall multiplicity is reduced due to energy transfer into the medium, we can thus understand the trend of the DRAG scenario. If a cutoff of 4 GeV is imposed, the trends qualitatively remain, with the exception of the RAD scenario which moves closer to the vacuum result. This again shows that the medium-induced changes in the RAD scenario chiefly affect the low $P_T$ part of the shower, which was noted before when studying the thrust distribution with cutoff imposed. These findings are qualitatively in agreement with the radiative energy loss scenario FMED as implemented in the MC code JEWEL \cite{JEWEL}.

Unlike other observables such as the nuclear suppression factor for single hadrons or the longitudinal momentum distribution in the jet studied previously \cite{YAS3}, thrust and especially $n$-jet fraction offer thus in principle the opportunity to distinguish details of the jet-medium interaction even on the level of distinguishing different implementations of radiative energy loss if they can be measured reliably in A-A collisions.

\section{Jet shapes}

\begin{figure*}[htb]
\begin{center}
\epsfig{file=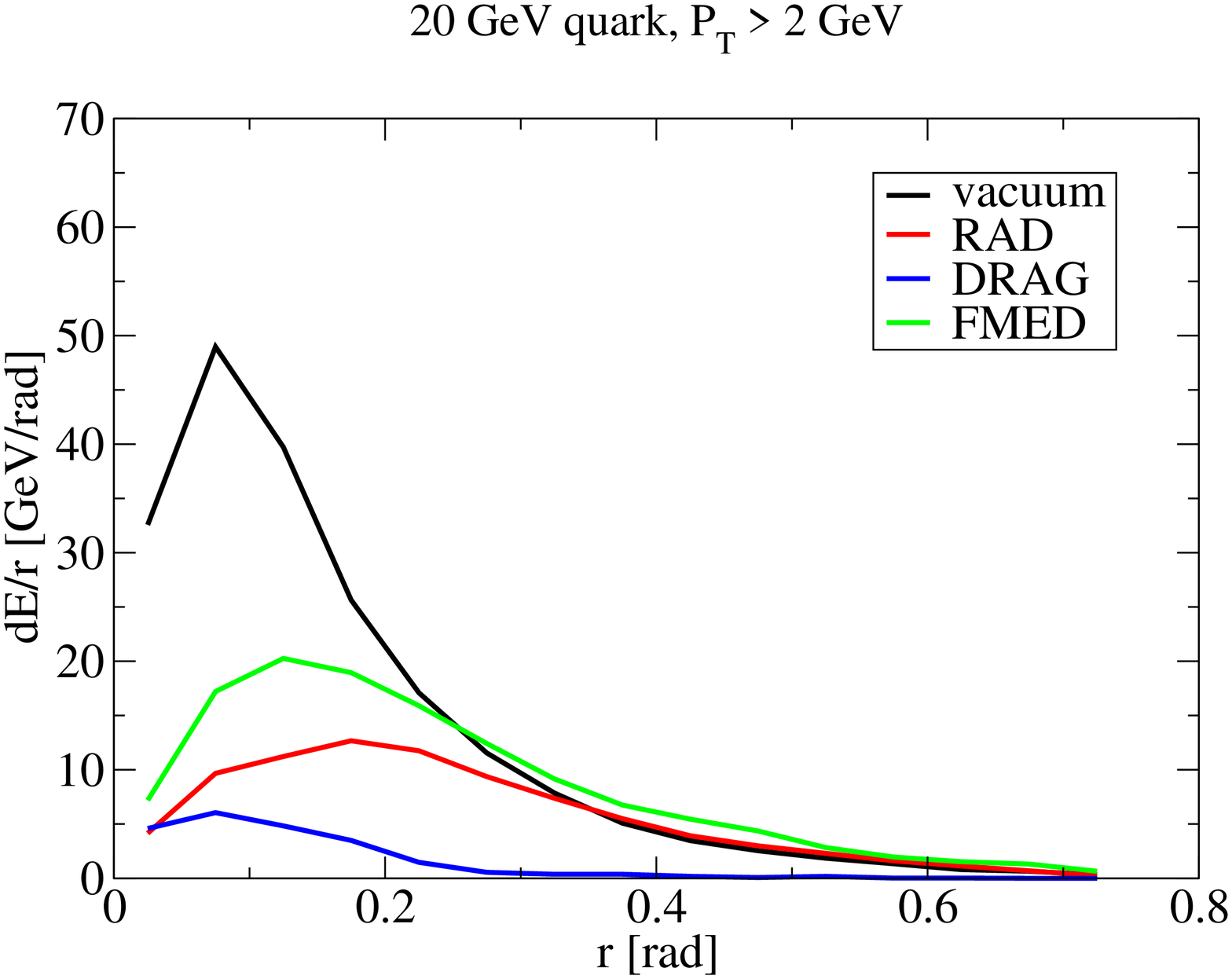,width=8cm}\epsfig{file=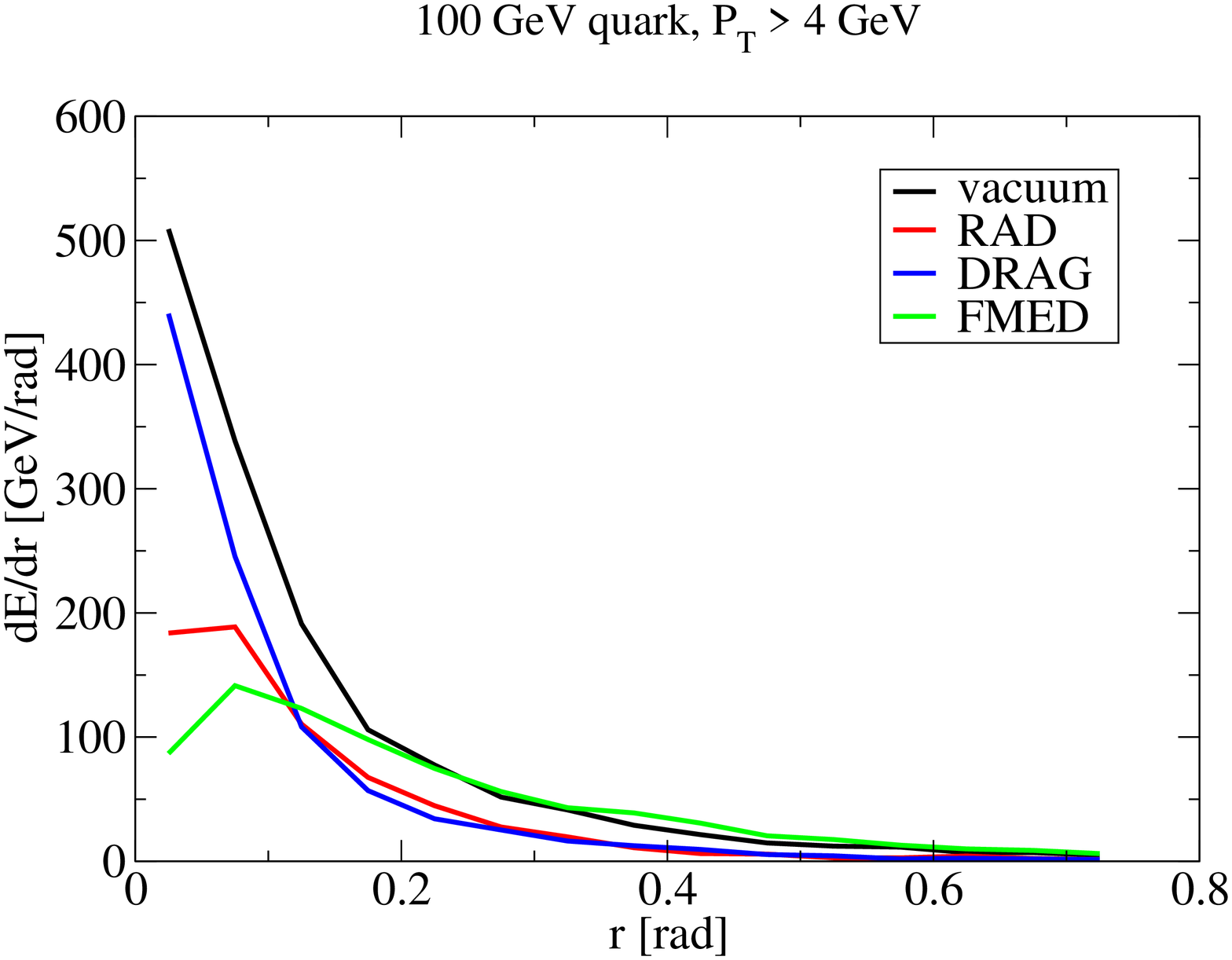, width=8cm}
\end{center}
\caption{\label{F-dEdr}(Color online) Differential energy flux as a function of angle with the jet axis $r$ for jets in vacuum and three different in-medium scenarios (see text) with a cutoff imposed for RHIC kinematics (left panel) and LHC kinematics (right panel). }
\end{figure*}

The jet shape measures the angular distribution of the flow of energy transverse to the jet axis. Once a jet axis and the particles $i$ associated with the jet have been identified experimentally, the integral jet shape given a cone radius $R$ is defined as

\begin{equation}
\Psi_{int}(r,R) = \frac{\sum_i E_i \theta(r-R_i)}{\sum_i E_i \theta(R-R_i)}
\end{equation}

where $E_i$ is the transverse energy of particle $i$, $r$ is an opening angle and $R_i$ is the angles of particle $i$ with the jet axis. If the jet axis is located at pseudorapidity $\eta$ and azimuth $\phi$, then $R_i = \sqrt{(\eta_i-\eta)^2+(\phi_i-\phi)^2}$. $\Psi_{int}(r,R)$ is thus the fraction of energy inside a sub-cone radius $r$ if the total energy is contained inside a cone of radius $R$.
From the integral jet shape, the differential jet shape can be obtained as

\begin{equation}
\psi(r,R) = \frac{d \Psi_{int}(r,R)}{dr}
\end{equation}

which is the angular density of energy flow in the jet. Since the jet shape depends on the jet energy as jets with larger energy are more collimated, one would like to study medium modifications of the jet shape for events in vacuum and in medium for the same energy of the shower-initiating parton. However, as mentioned before, since jet and medium interact, the energy of the shower-initiating parton is in general not the energy of the final state jet, and moreover one needs some criterion to even make a distinction between jet and medium in the soft sector. Thus, one needs a handle on the initial energy in order to make this comparison. Experimentally, this can be done e.g. using $\gamma$-jet correlations. 

In order to see that this is indeed a relevant effect and the energy of modified jets above the imposed cutoff is not roughly the same as the energy of the shower initiator, we show the angularly differential flux of energy $dE/dr$ for RHIC and LHC kinematics with the relevant medium cutoff imposed in Fig.~\ref{F-dEdr}. The difference in energy flux above the cutoff is more than a factor two between the different scenarios even at LHC kinematics. This strongly emphasizes the point that it is necessary to get a handle on the energy of the shower initiating parton.

\begin{figure*}[htb]
\begin{center}
\epsfig{file=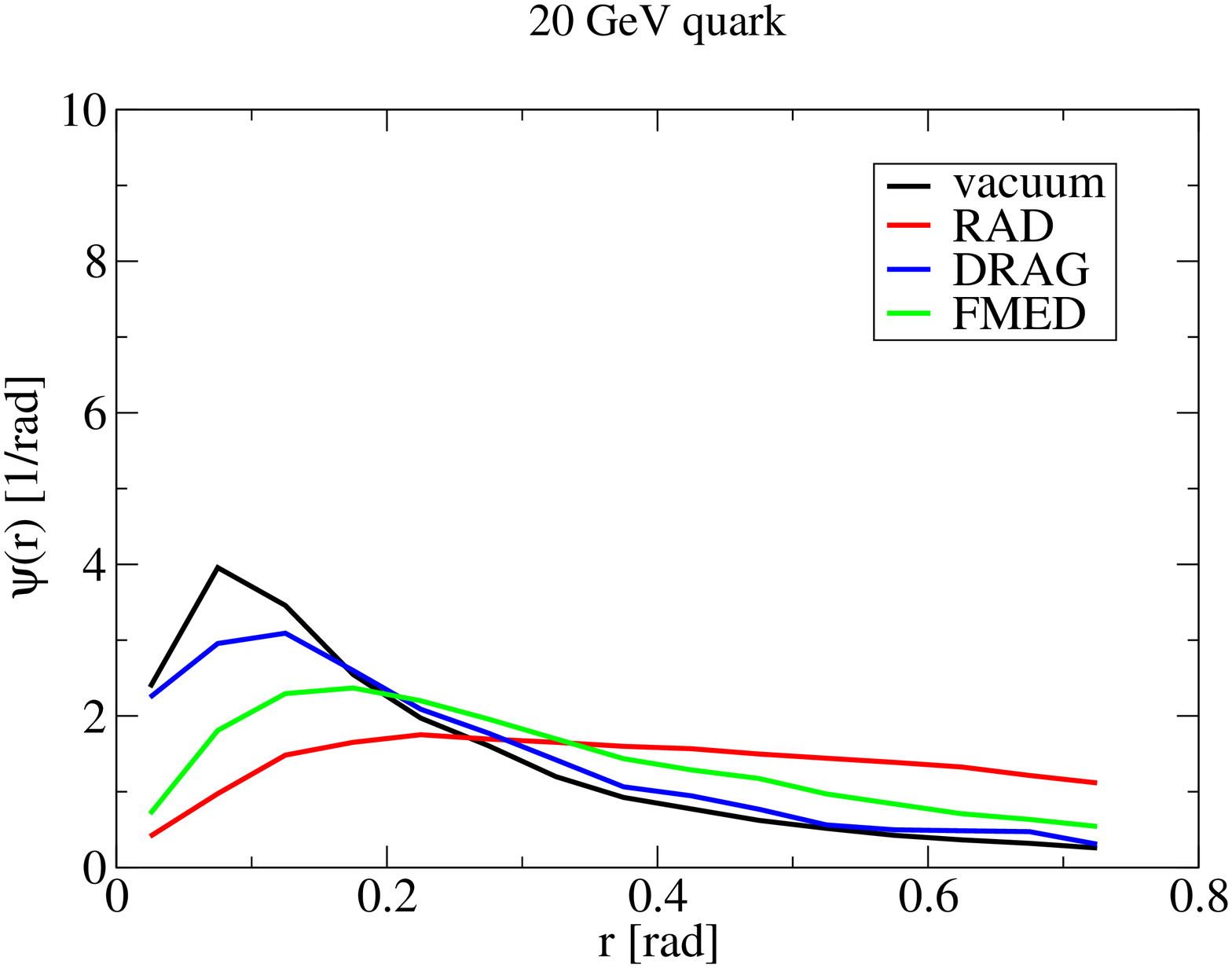,width=8cm}\epsfig{file=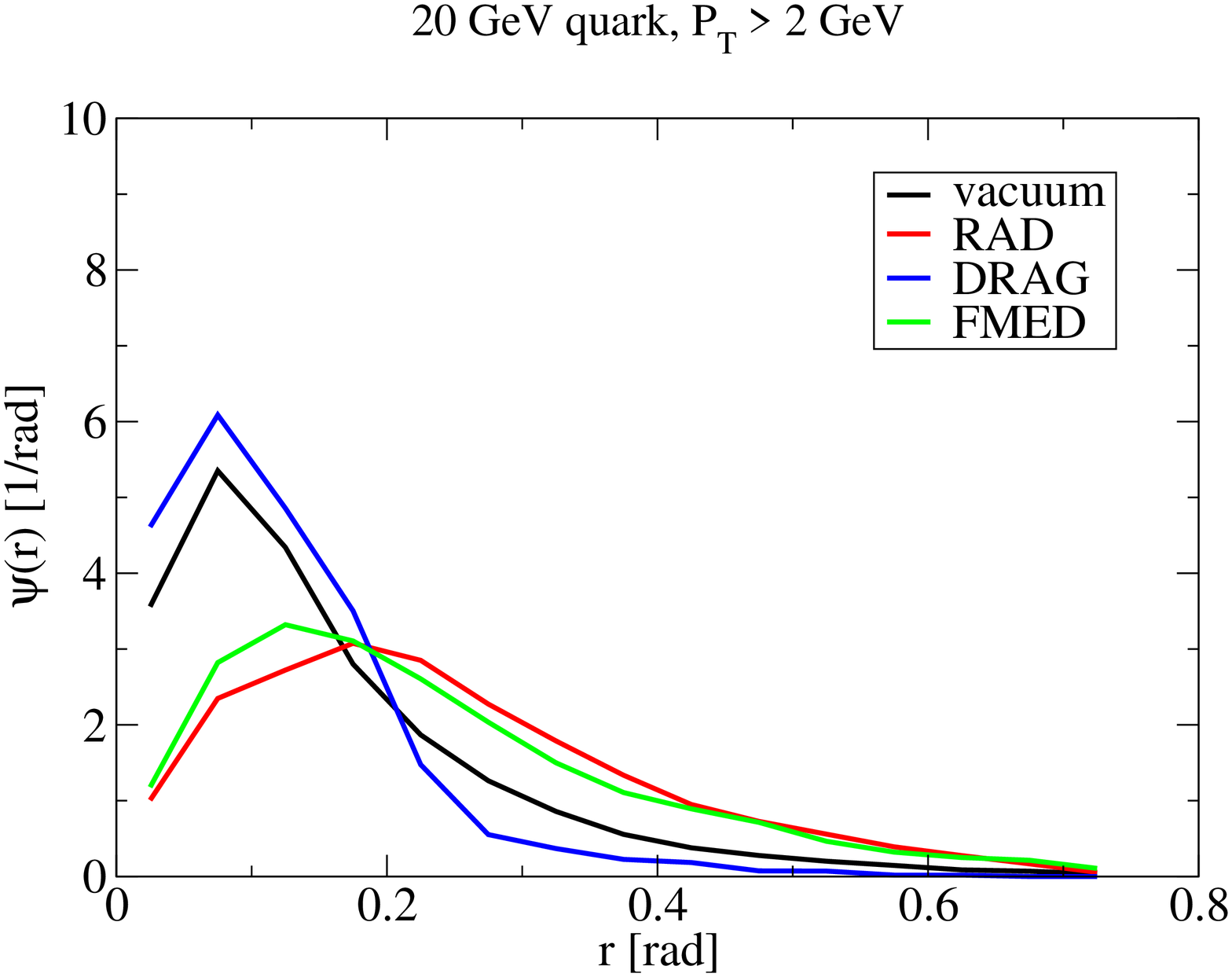, width=8cm}
\end{center}
\caption{\label{F-JetShape}(Color online) Differential jet shapes $\psi(r)$ as a function of angle $r$ given $R=0.7$ for a 20 GeV shower initiating quark in vacuum and for three different in-medium scenarios (see text) for all final state particles (left panel) and a cutoff of $P_T>2$ GeV imposed (right planel). }
\end{figure*}

\begin{figure*}[htb]
\begin{center}
\epsfig{file=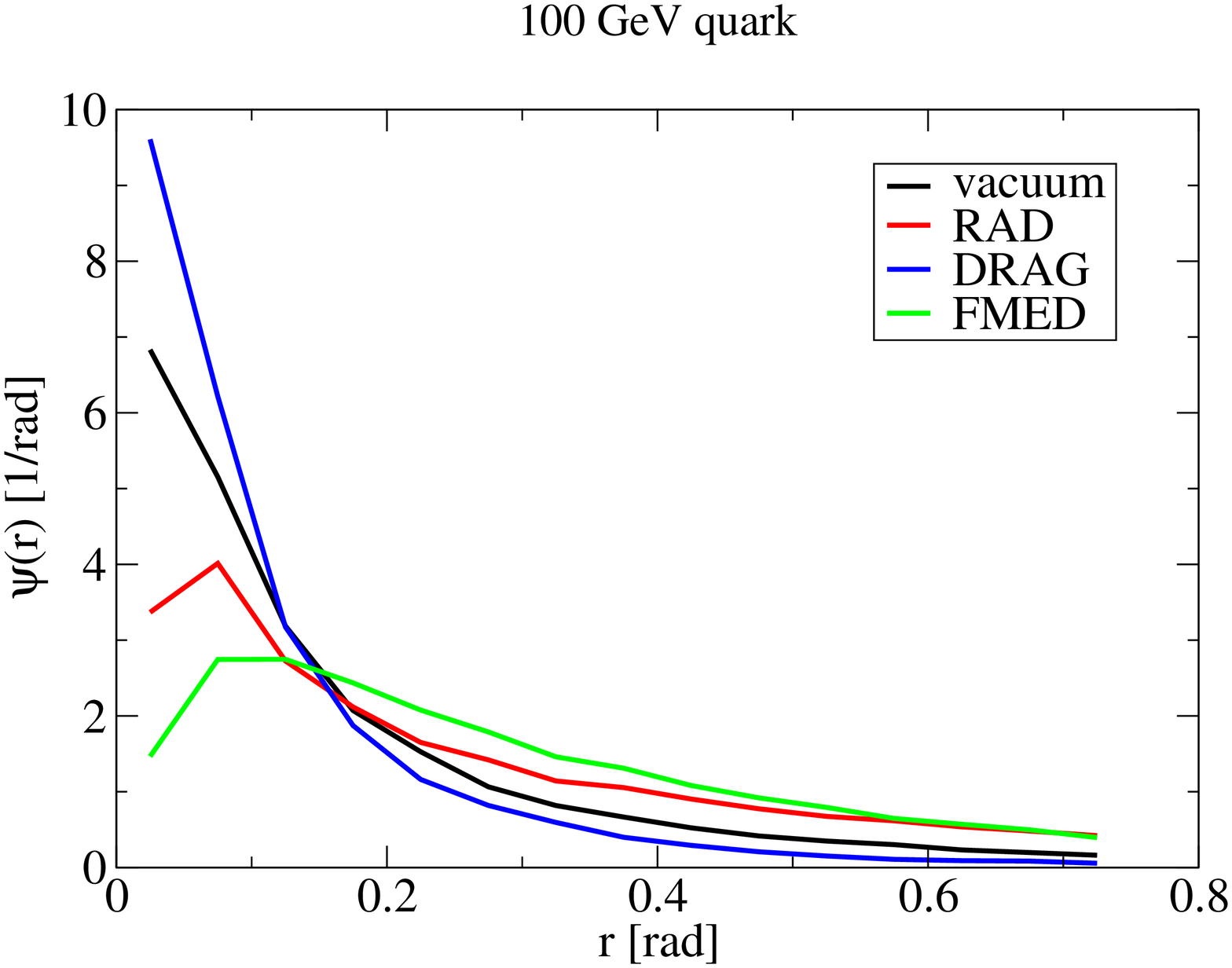,width=8cm}\epsfig{file=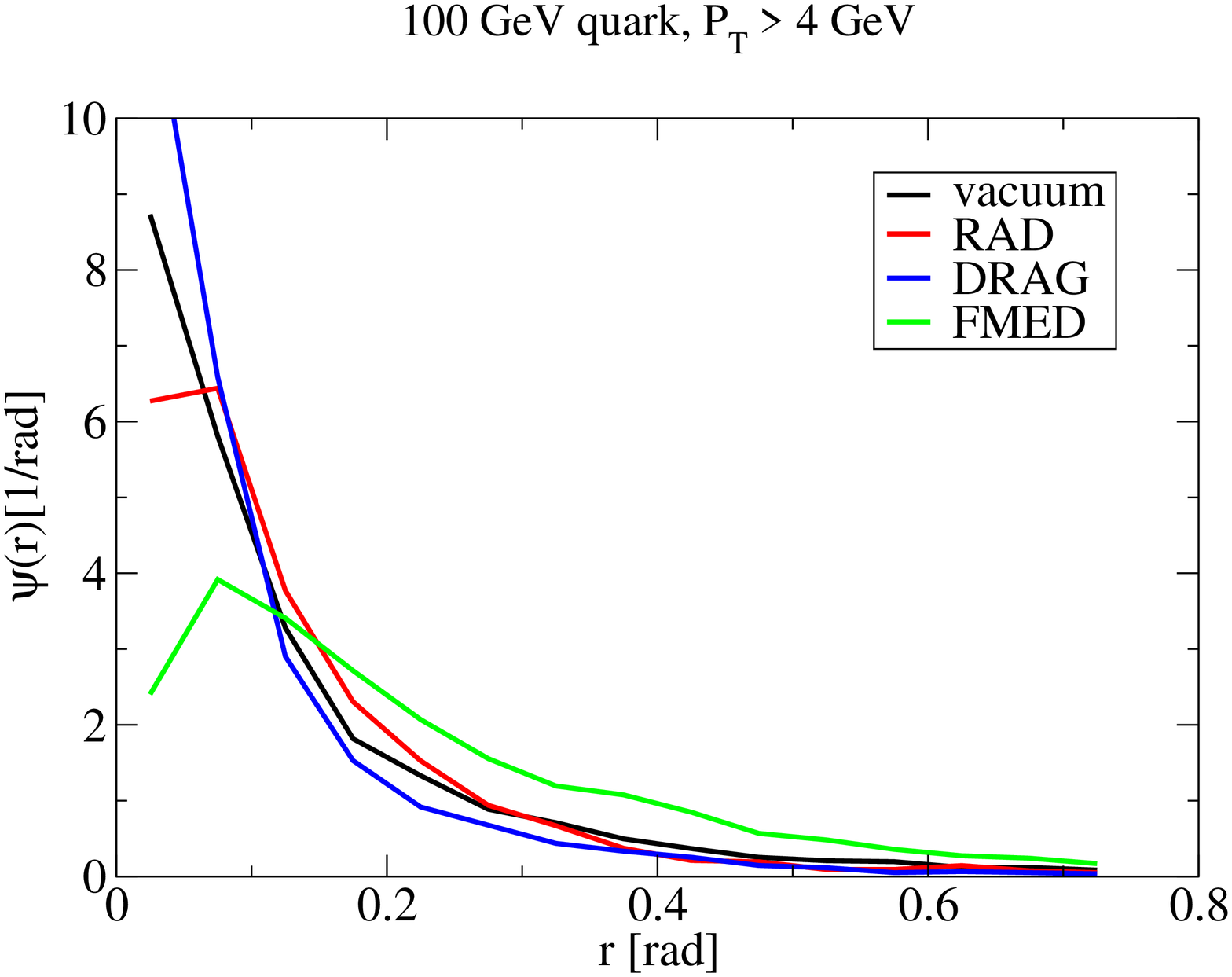, width=8cm}
\end{center}
\caption{\label{F-JetShapeLHC}(Color online) Differential jet shapes $\psi(r)$ as a function of angle $r$ given $R=0.7$ for a 100 GeV shower initiating quark in vacuum and for three different in-medium scenarios (see text) for all final state particles (left panel) and a cutoff of $P_T>4$ GeV imposed (right planel). }
\end{figure*}

Differential jet shapes for RHIC kinematics, i.e. a 20 GeV quark in vacuum and for the different in-medium scenarios are shown in Fig.~\ref{F-JetShape}, for LHC conditions the same is shown in Fig.~\ref{F-JetShapeLHC}. The observed picture is very consistent with what was found before: While the medium in the DRAG scenario tends to collimate the jet, the radiative scenarios RAD and FMED widen it and lead to energy flux at larger angles. However, there is an important difference between the RAD and the FMED scenario: In the RAD scenario large angle radiation appears only at low $P_T$, hence the observed widening is significantly reduced once a cutoff is imposed. This is not so in the FMED scenario where the effect persists despite a cutoff.

\section{Jet triggers and bias}

At least for LHC conditions, one may conclude from Figs.~\ref{F-dEdr} and \ref{F-JetShapeLHC} that on average most of the energy flux of the jet is within a relatively narrow cone $R<0.5$ and that even above a cutoff of $P_T>4$ GeV the integrated flux is substantial, regardless which in-medium scenario is studied. Thus, it should definitely be possible to directly trigger on medium-modified jets (an energy flux within a narrow cone) instead of using $\gamma$-jet coincidences.

The main problem with direct jet identification is to determine the total energy of the jet and to separate it from the background. There are three different contributions within a given angular region: 1) the (perturbative) jet itself, 2) medium which is (non-perturbatively) correlated with the jet due to the jet-medium interaction (e.g. through elastic recoil of medium particles, shockwaves, sound mode excitations in a strongly coupled fluid medium\dots) and 3) uncorrelated background medium. Since the average energy contained in the uncorrelated background can be observed event by event, and since due to energy conservation the sum of energy contained in the jet and the medium correlated with the jet must equal the energy of the shower-initiating parton, one might think of a strategy where a jet is identified above a cutoff in $P_T$ and its total energy is determined by subtracting the energy content of the uncorrelated background and to correct for the jet energy below the cutoff. This procedure requires that fluctuations around the average energy contained in the jet above the cutoff and in the background medium are small. 

Let us stress at this point again the fact that while we can formally calculate the production of soft partons in a medium-modified shower and find most of them still contained in a rather narrow cone (which would allow to get the total jet energy by making an angular cut without $P_T$ cut and just subtract the uncorrelated background), there is no reason to assume that soft partons or hadrons can be treated as part of a perturbative shower in the presence of a medium. On the contrary, good evidence has been found in two particle correlations \cite{PHENIX-2pc,STAR-2pc} and three particle correlations \cite{STAR-3pc} at RHIC energies that the distribution of soft hadrons correlated with a hard trigger is very different in vacuum and medium. In essence, energy and momentum flow through soft hadrons is observed at large angles. This has been variously interpreted as jet-induced shock waves in the medium \cite{Mach1,Mach2,Mach3,Mach4}. Given these finding, it is rather likely that a medium-modified jet is only a useful concept in terms of an energy flux in a relatively narrow angular region above a $P_T$ cut, but that the soft dynamics is very different in vacuum and in medium.

\begin{figure*}[htb]
\begin{center}
\epsfig{file=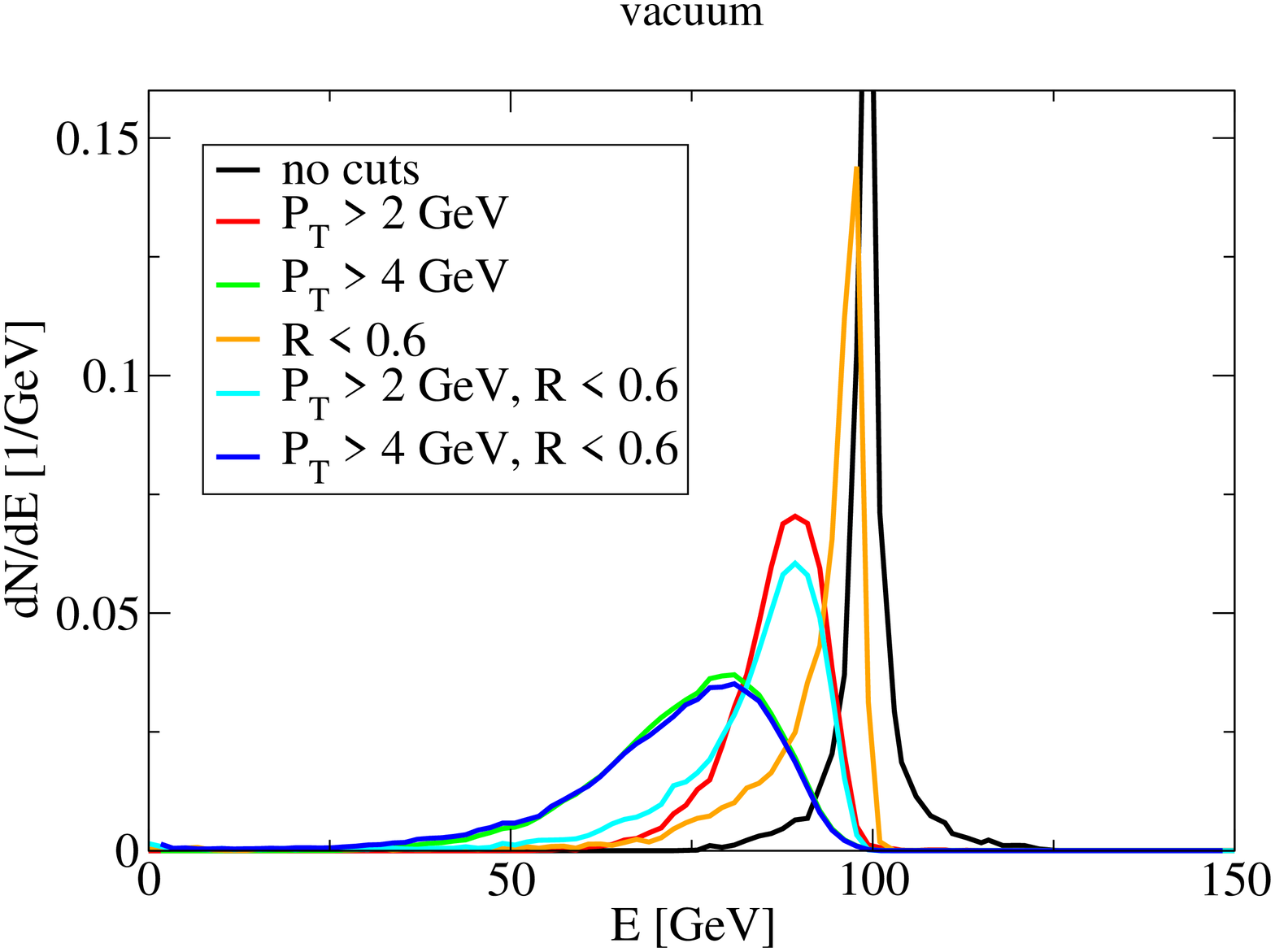, width=8cm}\epsfig{file=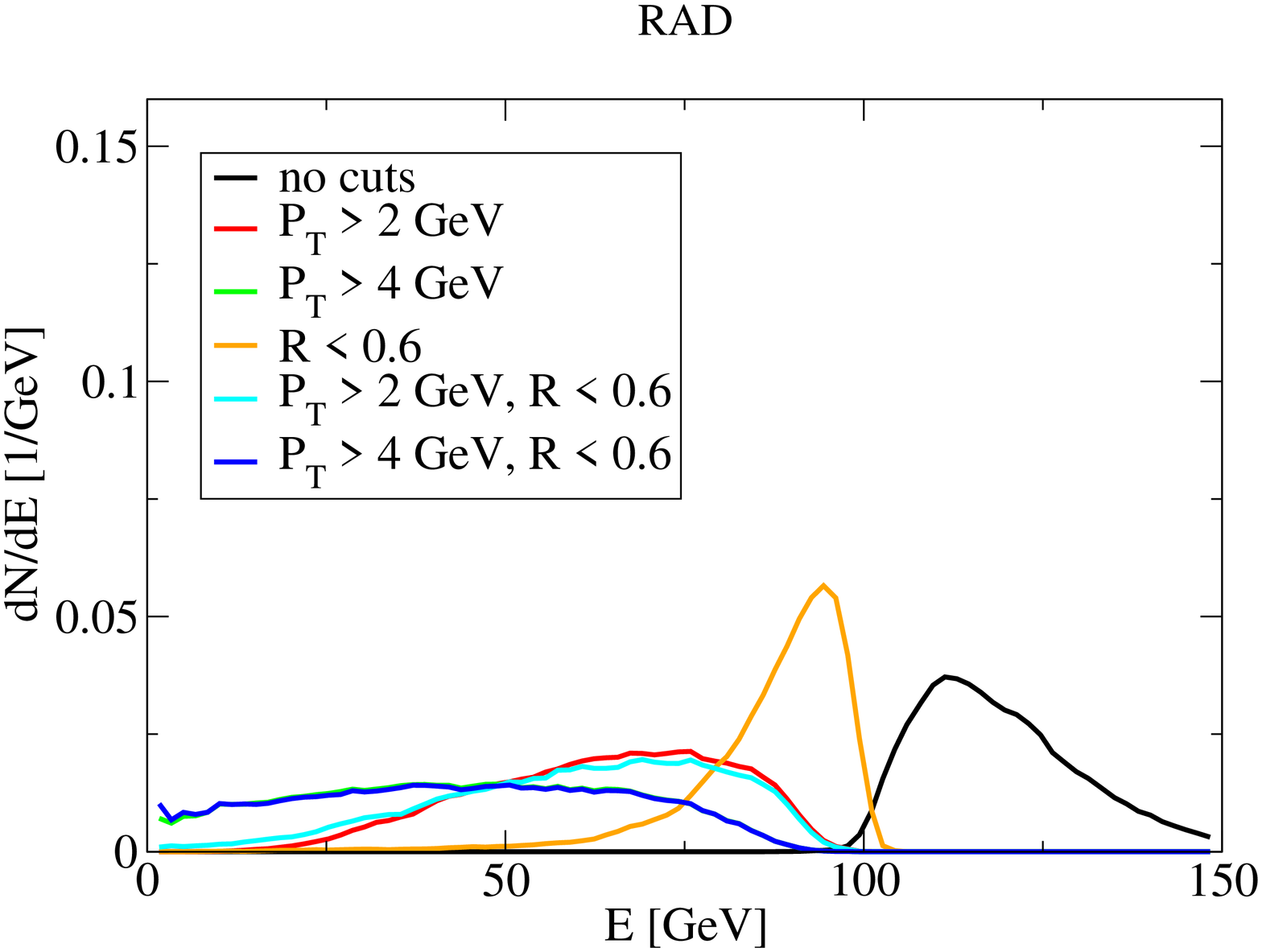, width=8cm}\\
\epsfig{file=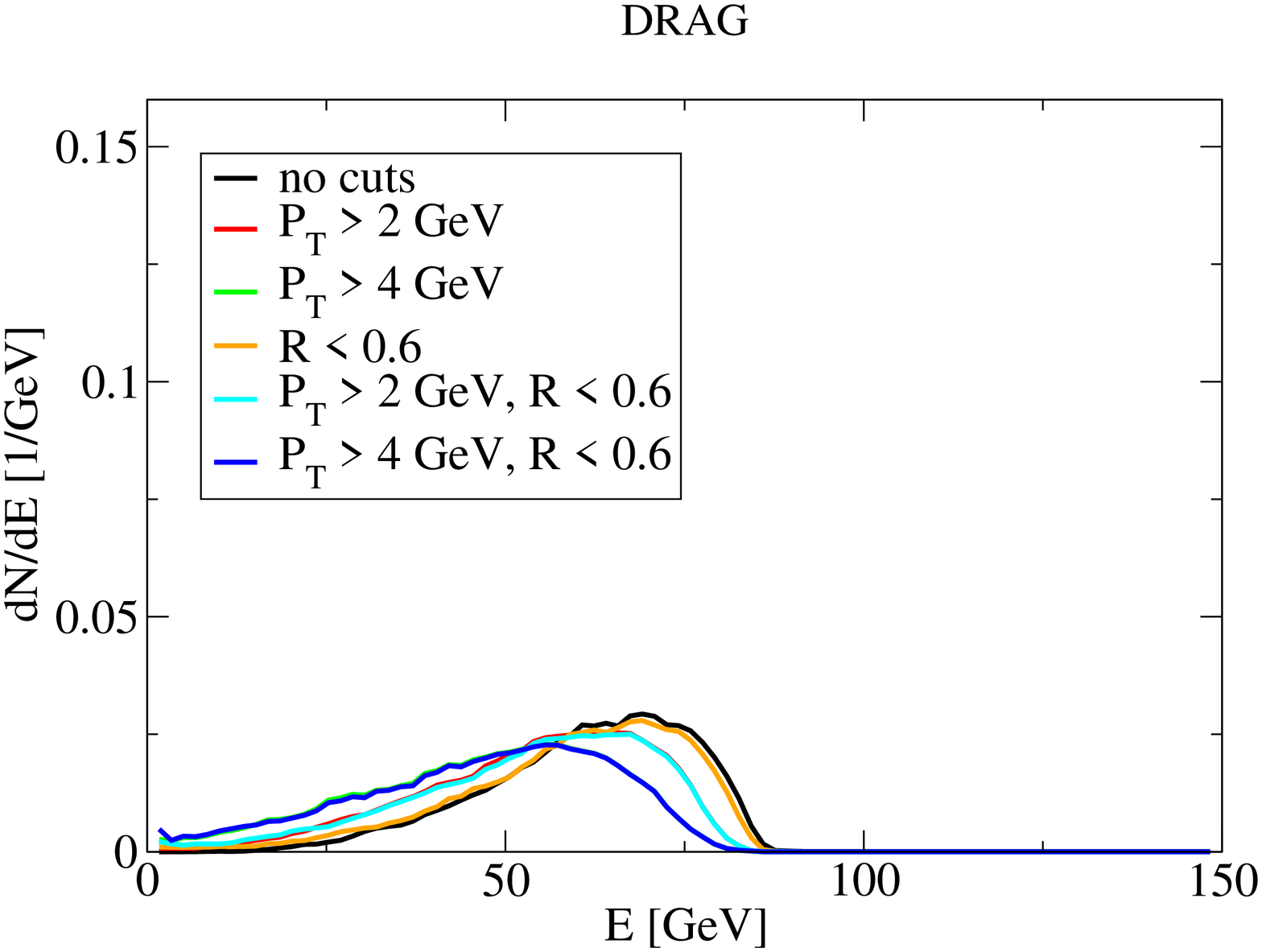, width=8cm}\epsfig{file=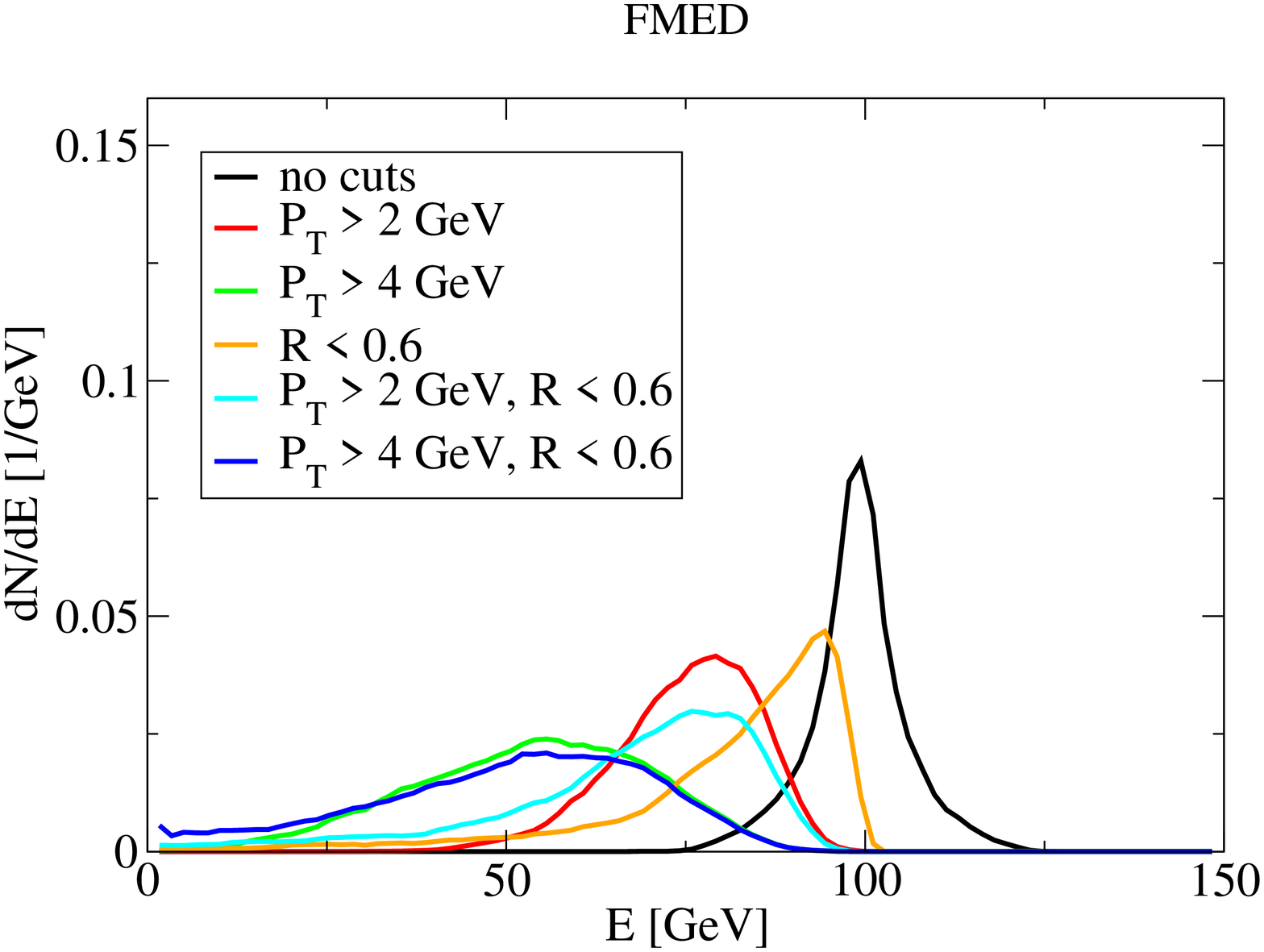, width=8cm}
\end{center}
\caption{\label{F-ETr}(Color online) Distribution of jets identified with energy $E$ for a given set of transverse momentum or angular cuts for a 100 GeV quark as shower initiator. Shown are vacuum (upper left) and the three different in-medium scenarios RAD (upper right), DRAG (lower left) and FMED (lower right).}
\end{figure*}

Thus, one needs to focus on the energy flux above a cutoff. If this is a quantity which does not fluctuate significantly, it may still be useful to estimate the total jet energy. It is beyond the scope of this paper to discuss fluctuations of the uncorrelated background, but it is rather easy within a MC simulation to discuss fluctuations of the energy within a given  set of cuts.

A useful quantity in this context is the distribution of events $dN/dE$ leading to a given energy $E$ observed inside the cuts, given that the jet was initiated by a parton with energy $E_0$. We show this quantity in Fig.~\ref{F-ETr} for a 100 GeV quark as showe initiator in vacuum and for the three different in-medium scenarios for an angular cut of $R<0.6$, the $P_T$ cut of 4 GeV used previously and a more optimistic $P_T$ cut of 2 GeV and the combination of the angular and the momentum cuts. Note that initial state effects as well as next to leading order pQCD diagrams lead to a momentum imbalance between two back-to-back partons. This can effectively be treated with a Gaussian distribution of  intrinsic $k_T$ which is added to the event, leading to the result that even in vacuum without any imposed cuts, the observed distribution of jet energy is never $\delta(E-E_0)$ but rather has a finite width.

It is evident from the figure that the medium leads to large fluctuations and that there is no good correlation between the energy detected inside the cuts and the initial energy of the jet. The fluctuations are especially large in the radiative energy loss scenarios: Even if a rather optimistic cut of $P_T>2$ GeV is imposed, the energy of the jet above the cut may be anything between 25 and 80 GeV in the RAD scenario. The situation is somewhat better in the FMED scenario, but even then the fluctuations are much increased as compared to the vacuum case. 

These results would suggest that it is very hard to get a reliable estimate for the initial energy of a medium-modified jet without resorting to $\gamma$-jet correlations. On the other hand, given a set of cuts in $R$ and $P_T$, there is a straightforward measure of the effect of energy flux outside the cut (although this cannot quantify the full probability distribution of finding energy outside the cuts) --- the nuclear suppression factor $R_{AA}$ of jets (see also the discussion in \cite{Shapes2}). If the cross section for producing a jet with energy $E$ within a set of cuts $R,P_T$ at rapidity $y$ is $\frac{d \sigma^{pp}}{dy dE}(E;R, P_T)$ and in heavy-ion collisions $\frac{d \sigma^{AA}}{dy dE}(E;R, P_T)$, then the nuclear suppression factor of jets is a straightforward generalization of the hadronic $R_{AA}$ as

\begin{equation}
R_{AA}^{jets}(E; R,P_T) = \frac{\frac{d \sigma^{AA}}{dy dE}(E;R, P_T)}{\langle N_{bin} \rangle \frac{d \sigma^{pp}}{dy dE}(E;R, P_T)}.
\end{equation}

In a similar way as hadronic $R_{AA} <1$ indicates that one is biased to observe high $P_T$ hadrons originating not from all events but only from those in which the medium effect was small can $R_{AA}^{jets} <1$ be used to study how biased the sample of jets identified within given cuts is. However, hadronic $R_{AA}$ cannot simply be unfolded to find the probability distribution of energy loss \cite{Inversion}, and for the same reasons jet $R_{AA}$ cannot in a straightforward manner be used to determine the probability distribution of energy deposition outside the cuts.

\section{Conclusions}

Identifying the precise nature of parton-medium interactions and hence also the relevant degrees of freedom in the medium is a major goal of the hard probes program within the physics of ultrarelativistic heavy-ion collisions. At present, different physics scenarios for this interaction, among them radiative energy loss, collisional energy loss or a medium-induced drag force are all (more or less) compatible with the suppression pattern of high $P_T$ hadrons as observed at RHIC (there are indications that the observed pathlength dependence in the suppression of back-to-back correlations rules out a significant amount of elastic energy loss though \cite{Elastic}). 

In this paper, it was shown that three such scenarios (two involving radiative energy loss, one a drag force) which lead to identical high $P_T$ hadron suppression at RHIC \cite{YAS3} could be potentially distinguished using jet observables such as the thrust distribution, the $n$-jet fraction or the jet shape. It should be stressed that even above a cutoff, the differences between the three scenarios appear substantial.

The major obstacle to a measurement however would be the precise quantification of the initial jet energy. In $\gamma$-jet correlation measurements, this is rather straightfoward, however such measurements are difficult and have limited statistics. In contrast, a direct identification of medium-modified jets and a correction for the energy flow outside the cuts would appear difficult as the event-by-event fluctuations of energy outside the cuts are significantly larger in the medium for all scenarios under investigation than in vacuum.  Note that the calculations presented here are also done for a fixed path. In a computation closer to the the experimental situation, the fluctuation of in-medium  pathlength also has to be taken into account. 

It is tempting to try to devise a suitable analysis strategy for jet measurements in A-A collisions at LHC from these findings. This, however, would in all likelihood be premature: How a medium-modified jet at LHC would look like is at present rather model dependent, but an optimized jet finding strategy would need to make use of just this knowledge. It seems much more promising to develop strategies for jet finding and analysis iteratively by contrasting first data with qualitative theoretical expectations and then refine both the theoretical framework and the experimental analysis strategy as needed. 

\begin{acknowledgments}
I'd like to thank Kari J. Eskola and J\"{o}rn Putschke for valuable discussions on the problem. This work was financially supported by the Academy of Finland, Project 115262. 
\end{acknowledgments}


\begin{thebibliography}{99}
\bibitem{Jet1}
  M.~Gyulassy and X.~N.~Wang,
  Nucl.\ Phys.\ B {\bf 420}, (1994) 583.
 
 
\bibitem{Jet2}
  R.~Baier, Y.~L.~Dokshitzer, A.~H.~Mueller, S.~Peigne and D.~Schiff,
  Nucl.\ Phys.\ B {\bf 484}, (1997) 265.
 
 
\bibitem{Jet3}
  B.~G.~Zakharov,
  JETP Lett.\  {\bf 65}, (1997) 615.
 
\bibitem{Jet4}
  U.~A.~Wiedemann,
  Nucl.\ Phys.\ B {\bf 588}, (2000) 303.
 
 
\bibitem{Jet5}
  M.~Gyulassy, P.~Levai and I.~Vitev,
  Nucl.\ Phys.\ B {\bf 594}, (2001) 371.
 
 
\bibitem{Jet6}
  X.~N.~Wang and X.~F.~Guo,
  Nucl.\ Phys.\ A {\bf 696}, (2001) 788.

\bibitem{PHENIX_R_AA}
  M.~Shimomura  [PHENIX Collaboration],
  nucl-ex/0510023.

\bibitem{Dijets1}
  D.~Magestro  [STAR Collaboration],
  nucl-ex/0510002; talk Quark Matter 2005.
 
\bibitem{Dijets2}
  J.~Adams {\it et al.}  [STAR Collaboration],
  nucl-ex/0604018.

\bibitem{PHENIX-RP}
  S.~S.~Adler {\it et al.}  [PHENIX Collaboration],
  Phys.\ Rev.\  C {\bf 76} (2007) 034904.

\bibitem{STARJET}
J.~Putschke  [STAR Collaboration],
  0809.1419 [nucl-ex].

\bibitem{HydroJet1}
  T.~Renk, J.~Ruppert, C.~Nonaka and S.~A.~Bass,
  Phys.\ Rev.\  C {\bf 75} (2007) 031902.


\bibitem{Dihadron1}
  T.~Renk,
  Phys.\ Rev.\  C {\bf 74} (2006) 024903.

\bibitem{Dihadron2}
  T.~Renk and K.~Eskola,
  Phys.\ Rev.\  C {\bf 75} (2007) 054910.

\bibitem{HydroJet2}
  A.~Majumder, C.~Nonaka and S.~A.~Bass,
  Phys.\ Rev.\  C {\bf 76} (2007) 041902.

\bibitem{HydroJet3}
  G.~Y.~Qin, J.~Ruppert, S.~Turbide, C.~Gale, C.~Nonaka and S.~A.~Bass,
  Phys.\ Rev.\  C {\bf 76} (2007) 064907.

\bibitem{Dihadron3}
  H.~Zhang, J.~F.~Owens, E.~Wang and X.~N.~Wang,
  Phys.\ Rev.\ Lett.\  {\bf 98} (2007) 212301.

\bibitem{JEWEL}
  K.~Zapp, G.~Ingelman, J.~Rathsman, J.~Stachel and U.~A.~Wiedemann,
  0804.3568 [hep-ph].

\bibitem{YAS}
T.~Renk,
   Phys.\ Rev.\  C {\bf 78} (2008) 034908.

\bibitem{YAS2}
 T.~Renk,
 0808.1803 [hep-ph].

\bibitem{YAS3}
 T.~Renk,
 0901.2818 [hep-ph].

\bibitem{Carlos}
  N.~Armesto, L.~Cunqueiro and C.~A.~Salgado,
  0809.4433 [hep-ph].

\bibitem{PYTHIA}
  T.~Sjostrand,
  Comput.\ Phys.\ Commun.\  {\bf 82} (1994) 74.


\bibitem{HERWIG}
  G.~Corcella {\it et al.},
  JHEP {\bf 0101} (2001) 010.

\bibitem{Shapes1}
  C.~A.~Salgado and U.~A.~Wiedemann,
  Phys.\ Rev.\ Lett.\  {\bf 93} (2004) 042301.

\bibitem{Shapes2}
  I.~Vitev, S.~Wicks and B.~W.~Zhang,
  JHEP {\bf 0811} (2008) 093.

\bibitem{ALEPH}
 A.~Heister {\it et al.}  [ALEPH Collaboration],
  Eur.\ Phys.\ J.\  C {\bf 35} (2004) 457.

\bibitem{PYSHOW}
  M.~Bengtsson and T.~Sj\"{o}strand, Phys.\ Lett.\ B {\bf 185} (1987) 435; Nucl. Phys.
B {\bf 289} (1987) 810; E.~Norrbin and T.~Sj\"{o}strand, Nucl.\ Phys.\ B {\bf 603} (2001) 297.

\bibitem{AdS}
 S.~S.~Gubser,
  Phys.\ Rev.\  D {\bf 74} (2006) 126005; C.~P.~Herzog, A.~Karch, P.~Kovtun, C.~Kozcaz and L.~G.~Yaffe,
  JHEP {\bf 0607} (2006) 013.

\bibitem{HBP}
  N.~Borghini and U.~A.~Wiedemann,
  hep-ph/0506218.

\bibitem{Opacity}
  S.~Wicks,
  0804.4704 [nucl-th].

\bibitem{Hydro}
  K.~J.~Eskola, H.~Honkanen, H.~Niemi, P.~V.~Ruuskanen and S.~S.~Rasanen,
  Phys.\ Rev.\ C {\bf 72} (2005) 044904.

\bibitem{Durham}
  S.~Catani, Y.~L.~Dokshitzer, M.~Olsson, G.~Turnock and B.~R.~Webber,
  Phys.\ Lett.\  B {\bf 269} (1991) 432.

\bibitem{PHENIX-2pc}
  S.~S.~Adler {\it et al.}  [PHENIX Collaboration],
  Phys.\ Rev.\ Lett.\  {\bf 97} (2006) 052301.

\bibitem{STAR-2pc}
J.~Adams {\it et al.}  [STAR Collaboration],
  Phys.\ Rev.\ Lett.\  {\bf 95} (2005) 152301.

\bibitem{STAR-3pc}
J.~G.~Ulery  [STAR Collaboration],
  Nucl.\ Phys.\  A {\bf 774} (2006) 581.

\bibitem{Mach1}
  H.~Stoecker,
  Nucl.\ Phys.\ A {\bf 750} (2005) 121.

\bibitem{Mach2}
  J.~Casalderrey-Solana, E.~V.~Shuryak and D.~Teaney,
  J.\ Phys.\ Conf.\ Ser.\  {\bf 27} (2005) 22
  [Nucl.\ Phys.\ A {\bf 774} (2006) 577].
 
\bibitem{Mach3}
 J.~Ruppert and B.~M{\"u}ller,
  Phys.\ Lett.\ B {\bf 618}, (2005) 123.

\bibitem{Mach4}
  T.~Renk and J.~Ruppert,
  Phys.\ Rev.\ C {\bf 73} (2006) 011901.

\bibitem{Inversion}
  T.~Renk,
  Phys.\ Rev.\  C {\bf 77} (2008) 017901.

\bibitem{Elastic}
  T.~Renk,
  Phys.\ Rev.\  C {\bf 76} (2007) 064905.

\end{thebibliography}
\end{document}